\documentclass[a4paper,fleqn,usenatbib]{mnras}

\usepackage{newtxtext,newtxmath}

\usepackage[T1]{fontenc}
\usepackage{ae,aecompl}

\usepackage[dvipdfmx]{graphicx}	
\usepackage{amsmath}	
\usepackage{amssymb}	
\usepackage{indentfirst}	
\usepackage{color}
\usepackage{pict2e}
\usepackage{bm}

\newcommand{\mr}{\mathrm} 

\newcommand{\msun}{\mr{M}_\odot} 
\newcommand{\zsun}{\mr{Z}_\odot} 

\newcommand{\mbh}{M_\mr{BH}} 
\newcommand{\ninf}{n_\infty} 
\newcommand{\rhoinf}{\rho_\infty} 
\newcommand{\tinf}{T_\infty} 
\newcommand{\cinf}{c_\mr{s,\infty}} 
\newcommand{\vflow}{v_\mr{flow}} 
\newcommand{\mach}{\mathcal{M}} 

\newcommand{\mbhf}{M_\mr{BH,4}} 
\newcommand{\ninfs}{n_{\infty,6}} 

\newcommand{\mbzero}{\dot{M}_\mr{B,0}} 
\newcommand{\rbzero}{R_\mr{B,0}} 
\newcommand{\mb}{\dot{M}_\mr{B}} 
\newcommand{\rb}{R_\mr{B}} 
\newcommand{\fdfb}{F_\mr{DF,B}} 

\newcommand{\fram}{F_\mr{ram}} 
\newcommand{\ftherm}{F_\mr{therm}} 
\newcommand{\dsh}{D_\mr{sh}} 

\newcommand{\rin}{R_\mr{in}} 
\newcommand{\rout}{R_\mr{out}} 

\newcommand{\ay}{a_\mr{Y}} 

\newcommand{\ledd}{L_\mr{E}} 
\newcommand{\medd}{\dot{M}_\mr{E}} 
\newcommand{\ledduv}{L_\mr{E,UV}} 
\newcommand{\medduv}{\dot{M}_\mr{E,UV}} 
\newcommand{\rhii}{R_\mr{HII}} 
\newcommand{\thii}{T_\mr{HII}} 
\newcommand{\nhii}{n_\mr{HII}} 
\newcommand{\chii}{c_\mr{s,HII}} 

\newcommand{\hii}{H{\sc ii} }

\newcommand{\vr}{v_\mr{R,crit}} 

\defcitealias{Park2017}{PB17}

\title[Dynamical friction onto moving IMBHs]
{Gaseous dynamical friction under radiative feedback: \\
do intermediate-mass black holes speed up or down?}

\author[Toyouchi et al.]{
Daisuke~Toyouchi$^1$, 
Takashi~Hosokawa$^1$, 
Kazuyuki~Sugimura$^2$, 
Rolf~Kuiper$^3$ \\
\\
$^{1}$Theoretical Astrophysics Group, Department of Physics, Kyoto University, Sakyo-ku, Kyoto 606-8502, Japan \\
$^{2}$Department of Astronomy, University of Maryland, College Park, MD 20740, USA \\
$^{3}$Institute of Astronomy and Astrophysics, University of T\"ubingen, Auf der Morgenstelle 10, D-72076 T\"ubingen, Germany
}

\date{Accepted XXX. Received YYY; in original form ZZZ}

\pubyear{2019}

\begin{document}
\label{firstpage}
\pagerange{\pageref{firstpage}--\pageref{lastpage}}
\maketitle

\begin{abstract}

Coalescence of intermediate-mass black holes (IMBHs) as a result of the migration toward galactic centers via dynamical friction may contribute to the formation of supermassive BHs. Here we reinvestigate the gaseous dynamical friction, which was claimed to be inefficient with radiative feedback from BHs in literature, by performing 3D radiation-hydrodynamics simulations that solve the flow structure in the vicinity of BHs. We consider a $10^4~\msun$ BH moving at the velocity $\vflow$ through the homogeneous medium with metallicity $Z$ in the range of $0-0.1~\zsun$ and density $\ninf$. We show that, if $\ninf \lesssim 10^{6}~\mr{cm^{-3}}$ and $\vflow \lesssim 60~\mr{km~s^{-1}}$, the BH is accelerated forward because of the gravitational pull from a dense shell ahead of an ionized bubble around the BH, regardless of the value of $Z$. If $\ninf \gtrsim 10^{6}~\mr{cm^{-3}}$, however, our simulation shows the opposite result. The ionized bubble and associating shell temporarily appear, but immediately go downstream with significant ram pressure of the flow. They eventually converge into a massive downstream wake, which gravitationally drags the BH backward. The BH decelerates over the timescale of $\sim 0.01$~Myr, much shorter than the dynamical timescale in galactic disks. Our results suggest that IMBHs that encounter the dense clouds rapidly migrate toward galactic centers, where they possibly coalescence with others.

\end{abstract}


\section{INTRODUCTION} \label{sec:intro}

The existence of supermassive black holes (SMBHs) exceeding $10^9~\msun$ at high redshifts (z $\gtrsim$ 6) is a big mystery in the modern astrophysics \citep[e.g.,][]{Fan2001, Willott2010, Mortlock2011, Venemans2013, Wu2015, Banados2018, Matsuoka2019}.
Intermediate-mass BHs (IMBHs) with $\mbh \sim 10^{3-5}~M_\odot$ are often considered as the seed objects \citep[e.g.,][]{Inayoshi2019}, and their coalescence should contribute to the rapid SMBH formation in the early universe \citep[e.g.,][]{Ryu2016,Tagawa2016}.
Once galaxies harboring IMBHs merge, the IMBHs are expected to drift in the remnant galaxy. 
The IMBHs gradually lose their orbital angular momentum due to dynamical friction from the surrounding stars or gas. If the frictional process operates efficiently, the IMBHs rapidly migrate inward and may form binaries. The IMBH binaries eventually coalescence yielding the gravitational wave emission, which is one of the main targets of the space-based interferometers such as LISA \citep[][]{LISA2013, LISA2017}.


The BH mass growth via IMBH coalescence is thought to be particularly important in the early universe. The latest cosmological galaxy formation simulations such as BlueTides \citep[e.g.,][]{Di_Matteo2017, Huang2019} show that the SMBH host galaxies have experienced numerous mergers with their surrounding galaxies. The resulting BH mergers drive the BH mass growth from $\mbh = 10^{3}~\msun$ to $\sim 10^{6}~\msun$ by the epoch of $z \simeq 12$. Such rapid growth is comparable to that via super-Eddington gas accretion, which does not necessarily occur \citep[e.g.,][]{Sugimura2018}. The BH mass further increases to $\sim 10^{9}~\msun$ via sub-Eddington accretion by $z \simeq 6$.


However, such cosmological simulations assume that BH coalescence occurs instantaneously after their host galaxies merge because the spatial resolution is not high enough to follow the BH orbital evolution. 
In order to clarify how efficiently the BHs migrate, BH orbital decay within each galaxy has to be solved. Indeed, \citet{Escala2005} and \citet{Mayer2007} perform such high-resolution simulations solving the temporal evolution of BH orbits in isolated galaxies. 
They show that the gaseous dynamical friction transports IMBHs from $\sim 100$~pc to sub-pc away from the centers just in $\sim 10$~Myr, which is much shorter than the galactic merger timescale. 
However, most of the previous studies have neglected effects of a copious amount of radiation emitted from the BHs. The resulting radiative feedback on the surrounding medium changes the efficiency of the dynamical friction. 
\cite{Souza-Lima2017} shows that the BH feedback is strong enough to disperse a gaseous wake downstream of the BH, the essential driver of the dynamical friction. Such a ``wake evacuation'' effect makes the BH orbital decay inefficient.
Park \& Bogdanovi\'c \citepalias[2017, hereafter][]{Park2017} have found a similar effect diminishing the gaseous friction on the IMBHs moving through the primordial gas. 
Their 2D radiation-hydrodynamic (RHD) simulations do not cover the galactic-scale gas dynamics, but spatially resolve the gas structure in the vicinity of the BHs, i.e., within the Bondi radius. 
A photoionized (or H{\sc ii}) bubble created around the BH destroys the downstream wake, which would appear without the radiative feedback. As a result the frictional force is substantially suppressed.


In this paper, we extend the previous study by \citetalias{Park2017} in the following two directions, to understand the gaseous friction exerted on IMBHs drifting in young galaxies. 
First, we consider the effects of heavy elements and dust grains contained in the gas through which the BHs migrate. A galaxy merger generally induces a starburst in the remnant galaxy, which scatters heavy elements and dust grains into the interstellar medium (ISM) \citep[e.g.,][]{Hopkins2008}. 
The gaseous dynamical friction on a moving BH is expected to be affected by the existence of dust grains because the size of the \hii bubble is reduced by the dust attenuation of the ultraviolet (UV) radiation \citep{Yajima2017, Toyouchi2019}.


Secondly, we explore very dense environments that have not been considered in \citetalias{Park2017}.
\cite{Inayoshi2016} study the gas accretion onto BHs with radiative feedback for such cases, assuming no relative velocities between the BH and ISM. They find that the flow structure qualitatively changes once the ambient density exceeds the threshold value $n_{\rm \infty, cr} \sim 10^6~\mr{cm^{-3}} (\mbh/10^4~\msun)^{-1}$.
The ram pressure of the accretion flow overcomes the thermal and radiation pressure within the bubble, and the \hii region is eventually trapped within the normal Bondi flow. \cite{Inayoshi2016} concludes that the resulting accretion rates are hardly affected by the radiative feedback and converge to the normal Bondi value. In our current study, we also consider the cases where the BH is moving through such dense environments. We expect that the density structure around BHs is qualitatively different from the case with $n_\infty \lesssim n_{\rm \infty, cr}$; the strong ram pressure of the flow should affect the density structure around the ionized bubble. We examine such a complex flow structure and the resulting gaseous dynamical friction by using the 3D RHD simulations.


The rest of the paper is organized as follows. We first summarize the underlying physics on the mass accretion and dynamical friction onto IMBHs in Section \ref{sec:basic}.
We next describe the method of our 3D RHD simulations in Section \ref{sec:method}.
The simulation results are given in Section \ref{sec:result}, based on which we discuss the actual orbital evolution of IMBHs in merged galaxies in Section \ref{sec:discussion}.
Finally, we provide a summary and conclusion in Section \ref{sec:summary}.


\section{UNDERLYING PHYSICS} \label{sec:basic}

In this section, we summarize the underlying physics which is useful to interpret our numerical results presented in Section~\ref{sec:result}. Consider a homogeneous flow with number density $\ninf$ and gas temperature $\tinf$ at the relative velocity $\vflow$ to an IMBH with mass $\mbh$. 
The Bondi radius, inside which the gravitational energy of the BH exceeds the gas thermal and kinetic energies, is given by
\begin{eqnarray}
\rb = \frac{G \mbh}{c^2_\mr{s,\infty}+v^2_\mr{flow}} = \frac{c^2_\mr{s,\infty}}{c^2_\mr{s,\infty}+v^2_\mr{flow}} \rbzero \ ,
\label{eq:rb}
\end{eqnarray}
\begin{eqnarray}
\rbzero = 1.4 \times 10^5~\mr{AU} \left ( \frac{\mbh}{10^4 \ \mr{M_\odot}} \right ) \left ( \frac{\tinf}{10^4 \ \mr{K}} \right )^{-1} \ ,
\label{eq:rbzero}
\end{eqnarray}
where $\cinf = \sqrt{k_\mr{B} \tinf / (\mu m_\mr{p})} = 8.1 (\tinf/10^4 \ \mr{K})^{1/2} \ \mr{km \ s^{-1}}$ is the sound speed for the isothermal gas. We first consider the so-called Bondi-Hoyle-Lyttleton (BHL) case, where radiative feedback from the BH accretion disk is neglected. In this case, the gas passing near the BH, particularly with the impact parameter less than the Bondi radius, is pulled back by the BH gravity to accumulate in the the backside of the BH. Such downstream structure gravitationally drags the BH backward, which is well known as the gaseous dynamical friction. \cite{Ostriker1999} analytically evaluates the frictional force and derives a formula for $\mach \equiv \vflow/\cinf > 1$ as,
\begin{eqnarray}
\fdfb = - \frac{4 \pi (G \mbh)^2 \mu m_\mr{p} \ninf}{v^2_\mr{flow}} \mr{ln} \left [ \Lambda \left ( 1 - \frac{1}{ \mach } \right ) \right ] \ ,
\label{eq:fdfb}
\end{eqnarray}
where $\mr{ln}\Lambda \equiv \mr{ln}\left(r_\mr{max}/r_\mr{min}\right)$ is Coulomb logarithm, and $r_\mr{max}$ and $r_\mr{min}$ are the maximum and minimum scales that describe the extent of the distribution of the gas contributing to the force. The accretion rate onto the BH is given by the BHL rate
\begin{eqnarray}
\mb = \frac{4 \pi \lambda_\mr{B} \mu m_\mr{p} \ninf (G \mbh)^2}{(c^2_{s,\infty}+v^2_\mr{flow})^{3/2}} = \left (\frac{c^2_{s,\infty}}{c^2_{s,\infty}+v^2_\mr{flow}} \right )^{3/2} \mbzero \ , 
\label{eq:mb}
\end{eqnarray}
\begin{eqnarray}
\mbzero = 1.7 \times 10^{-1} M_\odot \ \mr{yr}^{-1} \left ( \frac{\ninf}{10^5 \ \mr{cm}^{-3}} \right ) \left ( \frac{\mbh}{10^4 \ \mr{M_\odot}} \right )^2 \left ( \frac{\tinf}{10^4 \ \mr{K}} \right )^{-3/2} \ . \nonumber
\label{}
\end{eqnarray}
\begin{eqnarray}
\label{eq:mbzero}
\end{eqnarray}


Next we consider the effects of the radiation emitted from the BH accretion disk with $L = \eta \dot{M} c^2$, where $\dot{M}$, $\eta$, and $c$ are mass accretion rate, the radiative efficiency, and the speed of light, respectively.
The luminosity $L$ and corresponding accretion rate may be limited by the so-called Eddington values
\begin{eqnarray}
\ledd = \frac{4 \pi G \mbh c}{\kappa_\mr{T}} = 3.3 \times 10^8~L_\odot \left ( \frac{\mbh}{10^4 \ \mr{M_\odot}} \right ) \ ,
\label{eq:ledd}
\end{eqnarray}
\begin{eqnarray}
\medd &=& \frac{\ledd}{\eta c^2} \nonumber \\
&=& 2.2 \times 10^{-4}~M_\odot \ {\rm yr}^{-1} 
\left ( \frac{\mbh}{10^4 \ \mr{M_\odot}} \right )
\left( \frac{\eta}{0.1} \right)^{-1} \ ,
\label{eq:medd}
\end{eqnarray}
where $\kappa_\mr{T} = 0.4~\mr{cm^2~g^{-1}}$ is the opacity of Thomson scattering. The above limits implicitly assume that the accreting gas is fully ionized with the primordial composition. 
In our work, however, we consider the accretion flow containing dust grains.
Since the dust opacity can be larger than the Thomson scattering value, we may set the more stringent limits on the luminosity and mass accretion rate as 
\begin{eqnarray}
\ledduv &=& X_\mr{d,UV} \ledd \ ,
\label{eq:ledduv}
\end{eqnarray}
\begin{eqnarray}
\medduv &=& \frac{\ledduv}{\eta c^2} \ ,
\label{eq:medduv}
\end{eqnarray}
\begin{eqnarray}
X_\mr{d,UV} \equiv \frac{\kappa_\mr{T}}{\kappa_\mr{T} + \kappa_\mr{d,UV}} = \left \{ {1 + 7.1 \left (\frac{Z}{10^{-2} \ Z_\odot} \right )} \right \}^{-1} \ ,
\label{eq:kaiuv}
\end{eqnarray}
where $\kappa_\mr{d,UV}$ is the the dust absorption opacity for UV light, for which we assume $\kappa_\mr{d,UV} = 2.8 \times 10^2 (Z/Z_\odot)~\mr{cm^2~g^{-1}}$ \citep[e.g.,][]{Yajima2017}.
As is evident, for $Z \gtrsim 10^{-3}~\zsun$, $L_{\rm E,UV}$ becomes lower than the normal Eddington value $L_{\rm E}$. The radiative force exerted on dust grains predominantly sets the effective Eddington limit for such cases.

\begin{figure}
\begin{center}
\includegraphics[width=8cm]{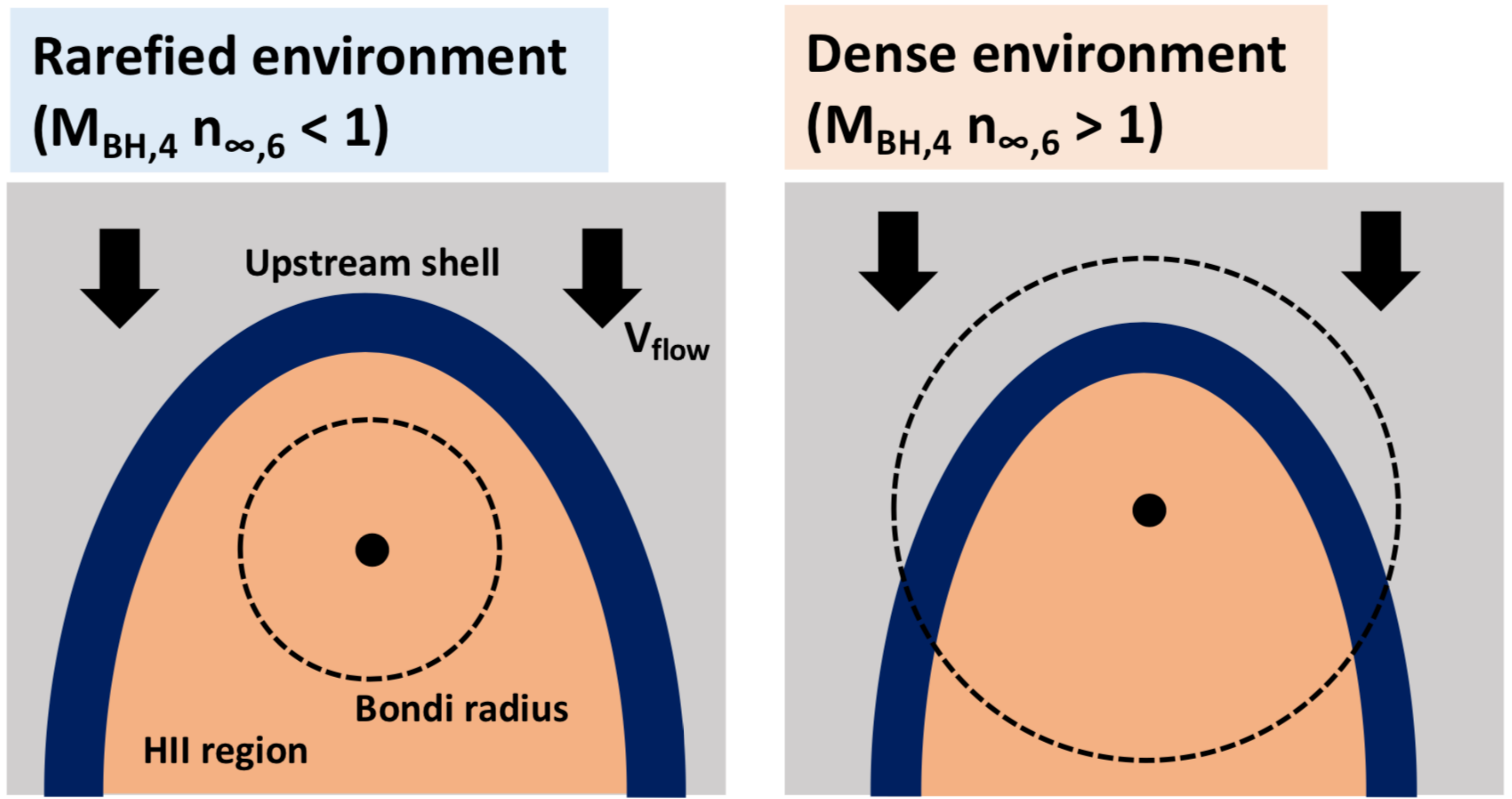}
\end{center}
\caption{Schematic pictures of the situations considered. The left and right pictures represent the cases where the ambient medium is rarefied and dense. The relative positions of the dense upstream shell around the \hii bubble and the Bondi radius are presented in each picture. 
}
\label{fig:dense_env}
\end{figure}

An ionized bubble created around the BH causes radiative feedback against the accretion flow.
When the BH is moving relative to the ISM, in particular, the \hii region has a cometary shape and a dense shell forms in the upstream side as illustrated in Figure \ref{fig:dense_env} \citep[e.g.,][\citetalias{Park2017}]{Park2013}. The downstream wake, which should appear in absence of the \hii bubble, significantly weaken or completely disappear. 
The BH is not efficiently decelerated for such a case, meaning that the radiative feedback substantially suppresses the gaseous dynamical friction.


However, we expect that the above picture should be modified if the effective Bondi radius is larger than the size of the \hii bubble, $\rhii$.
\cite{Inayoshi2016} investigate the rapid mass accretion onto BHs for cases with $\vflow = 0$.
According to them, for $\rb < \rhii$, an \hii bubble dynamically expands and disturbs the accretion flow toward the BH. Such a feedback effect substantially reduces the accretion rates from the Bondi value. For $\rb > \rhii$, by contrast, the \hii bubble collapses due to the ram pressure of the accretion flow. The accretion rates eventually converge to the Bondi value, for which the radiative feedback is of no effect.
In what follows, we investigate whether a similar transition occurs or not with non-zero $\vflow$ in the context of the gaseous dynamical friction.


For considering the ratio $\rb / \rhii$ with non-zero $\vflow$, we evaluate $\rhii$ as a distance from the BH to the dense upstream shell.
We suppose that the shell is in quasi-steady state where the momentum flux is conserved across the shell, namely $2 \nhii \chii^2 = \ninf (\cinf^2 + \vflow^2)$, where $\nhii$ and $\chii$ are gas number density and sound speed in \hii bubble, respectively.
We approximate $\rhii$ with the Str\"{o}mgren radius,
\begin{eqnarray}
\rhii &=& \left ( \frac{3Q_\mr{ion}}{4 \pi \alpha_\mr{rec,B} n^2_\mr{HII}} \right )^{1/3} \nonumber \\
&\propto& L^{1/3} T_\mr{HII} 
\left\{ \ninf (\cinf^2 + \vflow^2)  \right\}^{-2/3} \ ,
\label{eq:str_radius}
\end{eqnarray}
where $\alpha_\mr{rec} (\propto \thii^{-1})$ is the case-B hydrogen recombination coefficient, and $Q_\mr{ion} (\propto L)$ the emissivity of ionizing photons.
For acquiring the sufficient condition for $\rb > \rhii$, we evaluate the maximum size of the ionized bubble, which is realized with the Eddington luminosity given by Eq. (\ref{eq:ledduv}) as below,
\begin{eqnarray}
\rhii &=&  1.4 \times 10^5~\mr{AU} \left(1 + 7.1 \times \frac{Z}{10^{-2} \ Z_\odot} \right)^{-1/3}  \nonumber \\
& & \ \times \left ( \frac{\sqrt{\cinf^2+\vflow^2}}{10~\mr{km~s^{-1}}} \right )^{-4/3} \left ( \frac{\mbh}{10^4 \ M_\odot} \right )^{1/3} \left ( \frac{\ninf}{10^6 \ \mr{cm^{-3}}} \right )^{-2/3} \ , \nonumber \\
\label{eq:rhii}
\end{eqnarray}
where we assume the ionized gas temperature as $\thii = 7 \times 10^4~\mr{K}$, a typical value seen in our simulations. Although we here neglect the effect of dust attenuation of ionizing photons within the \hii bubble, this is a good approximation for our examined cases with $Z \lesssim 0.1~Z_\odot$ \citep[see Eq. 16 in][]{Toyouchi2019}.


After all, from Eqs. (\ref{eq:rb}) and (\ref{eq:rhii}), the condition of $\rb > \rhii$ for non-zero $\vflow$ is written as
\begin{eqnarray}
& & \left ( \frac{\mbh}{10^4 \ M_\odot} \right ) \left ( \frac{\ninf}{10^6 \ \mr{cm^{-3}}} \right ) \gtrsim  \ \ \ \ \ \ \ \ \ \ \ \ \ \ \ \ \ \ \ \ \ \ \ \ \ \ \ \ \ \ \ \ \ \ \ \ \ \ \ \ \ \ \ \ \ \ \ \ \  \nonumber \\
& &\ \ \ \ \ \ \ \ \ \ \ \ \ \ \ \ \ \ \ \    \left ( \frac{\sqrt{c^2_\mr{s,\infty}+v^2_\mr{flow}}}{10~\mr{km~s^{-1}}} \right ) \left(1 + 7.1 \frac{Z}{10^{-2} \ Z_\odot} \right)^{-1/2} \ . \nonumber \\
\label{eq:condi}
\end{eqnarray}
This equation implies that, for a fixed BH mass, dense environments are preferable to realize $\rb > \rhii$. We define the cases where Eq. (\ref{eq:condi}) is satisfied as the dense environments and the other cases as the rarefied environments (see also Figure \ref{fig:dense_env}).
We investigate the gas structure and the resulting BH acceleration realized in the rarefied and dense environments in Section \ref{sec:result}.
Since the right-hand side of Eq. (\ref{eq:condi}) is normally $O(1)$ in our examined cases with $Z < 0.1~Z_\odot$ and $\vflow \sim O(10)~\mr{km~s^{-1}}$, 
we approximately represent the condition of $\rb > \rhii$ as $\mbhf~\ninfs > 1$, where $\mbhf \equiv (\mbh/10^4~\msun)$ and $\ninfs \equiv (\ninf/10^6~\mr{cm^{-3}})$.

\section{SIMULATION METHOD} 
\label{sec:method}

We make use of a modified version of the public magneto-hydrodynamics code {\tt PLUTO} 4.1 \citep[e.g.,][]{Mignone2007}, which has been applied for studying the star formation at different metallicities \citep[e.g.,][]{Kuiper2010,Hosokawa2016,Nakatani2018a, Nakatani2018b, Kuiper2018, Kolligan2018}. In our previous studies, we have further implemented additional physics to investigate the gas accretion onto the BH under the radiative feedback \citep[][]{Sugimura2017, Sugimura2018, Toyouchi2019}.
We briefly describe our numerical method below \citep[see also][]{Toyouchi2019}.

\subsection{Basic setup} \label{sec:config}

We perform a suite of three-dimensional hydrodynamics simulations to solve the flow structure around the IMBHs. 
We use the 3D-spherical coordinates ($r, \theta, \phi$), in which a BH is located at the origin. 
Our computational domain covers the radial range from $\rin = 5 \times 10^3~\mr{AU} \simeq 1/30 \rbzero$ to $\rout = 10^7~\mr{AU} \simeq 70 \rbzero$, where $\rbzero$ is the Bondi radius for $\mbh = 10^4~\mr{M_\odot}$ and $\tinf = 10^4$ K.
We avoid too short time steps required to resolve the dense structure near the central BH accretion disk by using the sink region for $r \leq \rin$. 
The grid numbers are $(N_r, N_\theta, N_\phi) = (256, 36, 72)$ for all the examined cases. 
We adopt logarithmic spacing for the radial grids to realize the higher resolution in the inner region.

In this numerical domain, we consider the gas flow relative to the BH. 
We start our simulations assuming the homogeneous steady flow toward the negative Y-direction at the velocity $\vflow$ and solve the following governing equations,
\begin{eqnarray}
\frac{\partial \rho}{\partial t} + \nabla \cdot (\rho \bm{v}) = 0,  
\label{eq:mass_cons}
\end{eqnarray}
\begin{eqnarray}
\frac{\partial \rho v_r}{\partial t} + \nabla \cdot (\rho v_r \bm{v}) = - \frac{\partial P}{\partial r} + \rho \frac{v^2_\theta + v^2_\phi}{r} + \rho g_r \ ,  
\label{eq:mom_cons_r}
\end{eqnarray}
\begin{eqnarray}
\frac{\partial \rho v_\theta}{\partial t} + \nabla \cdot (\rho v_\theta \bm{v}) = - \frac{1}{r}\frac{\partial P}{\partial \theta} - \rho \frac{v_\theta v_r}{r} + \rho \frac{v^2_\phi~\mr{cot}~\theta}{r} + \rho g_\theta \ ,  
\label{eq:mom_cons_t}
\end{eqnarray}
\begin{eqnarray}
\frac{\partial \rho v_\phi}{\partial t} + \nabla \cdot (\rho v_\phi \bm{v}) = - \frac{1}{r~\mr{sin}~\theta}\frac{\partial P}{\partial \phi} - \rho \frac{v_\phi v_r}{r} - \rho \frac{v_\phi v_\theta~\mr{cot}~\theta}{r} + \rho g_\phi \ , \nonumber
\label{}
\end{eqnarray}
\begin{eqnarray}
\label{eq:mom_cons_p}
\end{eqnarray}
\begin{eqnarray}
\frac{\partial E}{\partial t} + \nabla \cdot (H \bm{v}) = \rho~\bm{v} \cdot \bm{g} + \rho~(\Gamma - \Lambda),  
\label{eq:ene_cons}
\end{eqnarray}
\begin{eqnarray}
\frac{\partial n_\mr{H} y_i}{\partial t} + \nabla \cdot (n_\mr{H} y_i \bm{v}) = n_\mr{H} R_i,  
\label{eq:chem_cons}
\end{eqnarray}
where $\rho$, $\bm{v} = (v_r, v_\theta, v_\phi)$, and $P$ are the gas density, velocity, and pressure, and $\bm{g} = (g_r, g_\theta, g_\phi)$ the external force including the BH gravity and radiative pressure, $E$ and $H$ the total energy and enthalpy per unit volume, $\Gamma$ and $\Lambda$ the specific heating and cooling rates. Magnetic fields and gas self-gravity are ignored in our simulations for simplicity. We discuss potential roles of the self-gravity later in Section \ref{sec:can_bh}.


With Eq. (\ref{eq:chem_cons}), we calculate the non-equilibrium chemical reactions for the eight species of HI, HII, HeI, HeII, HeIII, CII, OI, and $e^-$, where $y_i$ is the number ratio of $i$-th species to hydrogen nuclei and $R_i$ the corresponding chemical reaction rate. The CII and OI abundances are assumed to be constant at $y_\mr{CII} = 0.927 \times 10^{-4} Z/Z_\odot$ and $y_\mr{OI} = 3.568 \times 10^{-4} Z/Z_\odot$.
We also consider the dust grains contained in the gas, assuming the dust-to-gas mass ratio of $0.01 \times Z/Z_\odot$.
The reactions considered are photoionization and collisional ionization of HI, HeI and HeII,  and recombination of HII, HeII, HeIII. With the updated chemical abundances, we compute $\Lambda$ and $\Gamma$ summing up contributions from the photoelectric heating, fine-structure line cooling via CII and OI emission, free-free cooling of HI, HeI and HeII, and dust-gas collisional cooling.


In this study, we only examine the cases with $Z \leq 0.1~Z_\odot$, for which our minimal chemistry network provides reasonable approximations. Additionally, we ignore any molecular cooling, which could be important in $T \sim 10$ K. Note that in nearby galaxies the observed kinetic temperature of circum-nuclear disks (CNDs) is typically much higher than $\sim 10$~K \citep[e.g.,][]{Davies2012, Izumi2013, Viti2014}.

\subsection{Radiative Feedback} \label{sec:subgrid}

Mass accretion rates onto the unresolved accretion disk $\dot{M}$ are assumed to be given by the mass influx measured at $\rin$.
To incorporate the radiative feedback against the accretion flow, we evaluate the disk luminosity as functions of $\dot{M}$ using a sub-grid model,
\begin{eqnarray}
L =  
\begin{cases}
2 \ \ledd \ \left [ 1 + \mr{ln} \left ( \frac{\dot{m}}{2} \right ) \right ] & (\dot{m} > 2) \\
\ledd \ \dot{m} & (\mr{otherwise}) \ \ ,
\end{cases}
\label{eq:MtoL}
\end{eqnarray}
where $\dot{m}$ is defined as $\dot{M}/\medd$. The above formula well approximates the model results by \cite{Watarai2000}. Note that the luminosity $L$ does not largely surpass the Eddington value owing to the photon trapping effect.
We assume that the radiation is isotropically emitted from the sink, with a power-law spectrum $L_\nu \propto \nu^{-\alpha} \ (\alpha = 1.5)$ in a limited frequency range $6~{\rm eV} \leq h \nu \leq 1~{\rm keV}$.
We do not model winds or jets possibly launched from the accretion disk, assuming no outward mass flux from the sink. See Section \ref{sec:multi} for their potential effects on the accretion flows.


We solve the transfer of far-ultraviolet (FUV; 6 eV $\leq \nu \leq$ 13.6 eV) and extreme-ultraviolet (EUV; 13.6 eV $\leq \nu \leq$ 1 keV) photons emitted from the sink by means of the frequency-dependent ray-tracing method along the radial cells.
We consider the consumption of EUV photons by photoionization of HI, HeI, and HeII with the cross-sections given by \cite{Osterbrock1989} and \cite{Yan1998}, and the dust attenuation of FUV and EUV photons with the opacity table of \cite{Weingartner2001}. We also solve the transfer of diffuse IR photons coming from the thermal dust emission with the flux-limited diffusion (FLD) approximation method. We make use of the FLD module developed by \cite{Kuiper2010}, which has been well tested and used in a series of work \citep[e.g.,][]{Kuiper2011, Kuiper2012, Kuiper2018}.
We also compute the radiative force via Thomson scattering, photoionization, and dust absorption consistently with the obtained radiation fields. The further details on our radiative transfer method are also given in \cite{Toyouchi2019}.

\subsection{Calculation of BH acceleration} \label{sec:cal_bh_acc}

We calculate the BH acceleration in the Y-direction $\ay$ by summing up the gravitational force from the surrounding gas structure,
\begin{eqnarray}
\ay = \sum_i \frac{\mr{G} \rho_i \mr{sin}\theta_i \mr{sin}\phi_i}{r^2_i} \mr{d}V_i \ ,
\label{eq:ay}
\end{eqnarray}
where $\rho_i$ and $\mr{d}V_i$ are the mass density and volume of $i$-th cell located at ($r_i$, $\theta_i$, $\phi_i$).
Here, a positive (negative) value of $\ay$ means that the IMBH accelerates in the upstream (downstream) direction, i.e., the BH speeds up (down).


We note that the BH acceleration is calculated as a post-process. Namely, we fix $v_{\rm flow}$ during each run. We normally calculate the acceleration with a snapshot at $t = 0.4$~Myr, by which the flow reaches a quasi-steady state. Our treatment is valid because the resulting acceleration occurs over the timescale longer than that of our simulations. The actual time evolution of the BH velocity is discussed with $a_{\rm Y}$ obtained for different $v_{\rm flow}$ in Section \ref{sec:can_bh}.

\subsection{Cases examined} \label{sec:setup}

In our simulations, the BH mass is fixed at $\mbh = 10^4~\mr{M_\odot}$, and the incident flow density $\ninf$ and metallicity $Z$ are varied among models (Table~\ref{table:model}).  
The model name represents the values of $Z$, $\ninf$, and $\vflow$; for instance, our fiducial model Z2N4V20 represents $Z = 10^{-2}~Z_\odot$, $\ninf = 10^4~\mr{cm^{-3}}$, and $\vflow = 20~\mr{km~s^{-1}}$, with which the condition of $\mbhf~\ninfs < 1$ is satisfied.
The fiducial parameters of $\ninf = 10^4~\mr{cm^{-3}}$ and $\vflow = 20~\mr{km~s^{-1}}$ are chosen in reference to numerical simulations that follow the Milky Way-size galaxy mergers \citep[e.g.,][]{Mayer2007, Mayer2010, Roskar2015}. 
We set the gas temperature at the equilibrium value for given $\ninf$, $Z$, and the presumed background FUV and X-ray radiation fields. We assume the background fields 100 times stronger than in the solar neighborhood \citep[see][for more details]{Toyouchi2019}.


Apart from model Z2N4V20, we also consider models PRN4V20 and Z1N4V20, where only the metallicity is varied as $Z = 0$ and $0.1~Z_\odot$ to investigate the $Z$-dependence.
We also consider Z2N6V20 model with the higher density $\ninf = 10^6~\mr{cm^{-3}}$,  with which the condition $\mbhf~\ninfs > 1$ is satisfied. Such a dense environment corresponds to rare dense parts of the gas disk formed in merger remnant galaxies \citep[][]{Fiacconi2013, Souza-Lima2017}.


In addition to the above cases with $\vflow = 20~\mr{km~s^{-1}}$, we also investigate model Z2N4V100 with $\vflow = 100~\mr{km~s^{-1}}$. Such a high velocity is also observed in numerical simulations, as upper values of the rotational or turbulent velocities.
We note that, for Z2N4V100 model, we adopt the smaller inner boundary radius $\rin = 10^{2}~\mr{AU}$ as shown in Table \ref{table:model}, because the default sink size is bigger than the effective Bondi radius for $\vflow = 100~\mr{km~s^{-1}}$.

Incidentally, we performed two convergence check tests for the fiducial model. The one adopts the doubled number of the spatial grids, and the other the halved sink size. As a result, we confirmed good convergences in terms of the density structure and the resulting gaseous dynamical friction.


\begin{table*}
\begin{center}
\caption{Model parameters} \label{table:model}
\begin{tabular}{cccccccc} \hline \hline
Model                & $\mbh \ \rm [M_\odot]$ & $Z \ [\mr{Z_\odot}]$  & $\ninf \ \rm [cm^{-3}]$ & $\vflow \ \rm [km~s^{-1}]$ & $\tinf \ \rm [K]$ & $\rin \ \rm [AU]$ & $\rout \ \rm [AU]$ \\ \hline
PRN4V20           & $10^4$                         & $0$                          & $10^4$                        & $20$                                  & $6.8 \times 10^3$ & $5 \times 10^{3}$ & $10^{7}$ \\
Z2N4V20           & $10^4$                         & $0.01$                      & $10^4$                        & $20$                                  & $3.0 \times 10^2$ & $5 \times 10^{3}$ & $10^{7}$ \\
Z1N4V20           & $10^4$                         & $0.1$                        & $10^4$                        & $20$                                  & $1.8 \times 10^2$ & $5 \times 10^{3}$ & $10^{7}$ \\
Z2N4V100         & $10^4$                         & $0.01$                      & $10^4$                        & $100$                                & $3.0 \times 10^2$ & $10^{2}$               & $10^{7}$ \\
Z2N6V20           & $10^4$                         & $0.01$                      & $10^6$                        & $20$                                  & $8.9 \times 10$     & $5 \times 10^{3}$ & $10^{7}$ \\  \hline
\end{tabular}
\end{center}
\end{table*}


\section{RESULTS} \label{sec:result}

\subsection{Metallicity Dependence} \label{sec:metal}

\begin{figure*}
\begin{center}
\includegraphics[width=17cm]{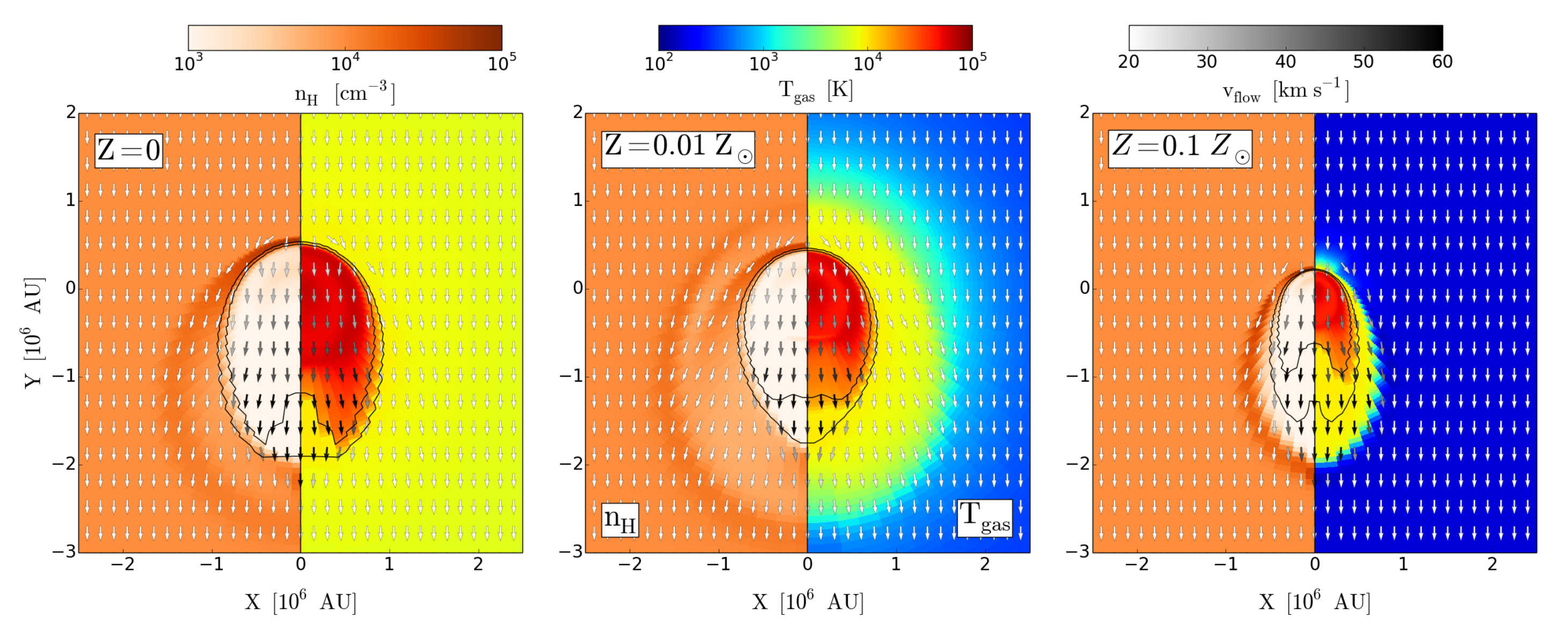}
\end{center}
\caption{
Metallicity-dependence of the gas structure around a moving and accreting IMBH. The panels show snapshots at $t = 0.4~\mr{Myr}$ in PRN4V20 ($Z = 0$), Z2N4V20 ($Z = 0.01~\zsun$), and Z1N4V20 ($Z = 0.1~\zsun$) models from left to right. Each panel presents the spatial distributions of the gas number density (left) and temperature (right) around the BH in an XY slice. The arrows represent the gas velocity vectors at different points. The inner and outer contours indicate the positions where the neutral hydrogen fraction is 0.01 and 0.99. The layer between these contours corresponds to the ionization front.
}
\label{fig:str_M4N4M2}
\end{figure*}

First, we investigate the metallicity-dependence of the flow structure around a moving BH. Figure \ref{fig:str_M4N4M2} shows the distributions of the gas number density and temperature in the quasi-steady states for the different metallicities of $Z = 0, ~0.01~\zsun$, and $0.1~\zsun$ (models PRN4V20, Z2N4V20, and Z1N4V20). We see, for the case with $Z=0$, an ionized bubble elongated downstream and a dense shell in the upstream side of the bubble. Such features are generally in agreement with the previous results obtained by 2D simulations for the similar primordial cases \citep[][\citetalias{Park2017}]{Park2013}.
The thermal structure for $Z = 0.01~\zsun$ is somewhat more complicated than for $Z = 0$. It consists of distinct cold and warm ($T \sim 10^2$ and $10^4$ K) layers around the hot ionized bubble. 
In addition to the upstream shell, an egg-like shell appears along the boundary between the cold and warm layers.
The cold region corresponds to the ambient gas in the thermal equilibrium state achieved by the balance between metal-line cooling and heating by the background radiation. 
The warm region is partly ionized by X-ray photons leaking from the ionized bubble, and the temperature is set by the balance between photoionization heating and Ly$\alpha$ cooling \citep[][]{Takeo2019}.
Such a double-shell structure disappears for the higher metallicity $Z = 0.1~\zsun$.
In this case, the metal cooling is so efficient that the warm region does not extend widely, and as a result, a sharp transition between the cold and hot media occurs just in front of the upstream ionization front.
The kinetic flow structure is common among these cases. The flow velocity gradually increases after crossing the dense upstream shell, from $v_{\rm flow} \simeq 20~ \mr{km~s^{-1}}$ to $\simeq 50~ \mr{km~s^{-1}}$. Such acceleration occurs because the flow thermal pressure increases across the ionization front owing to the photoionization heating.

\begin{figure}
\begin{center}
\includegraphics[width=8cm]{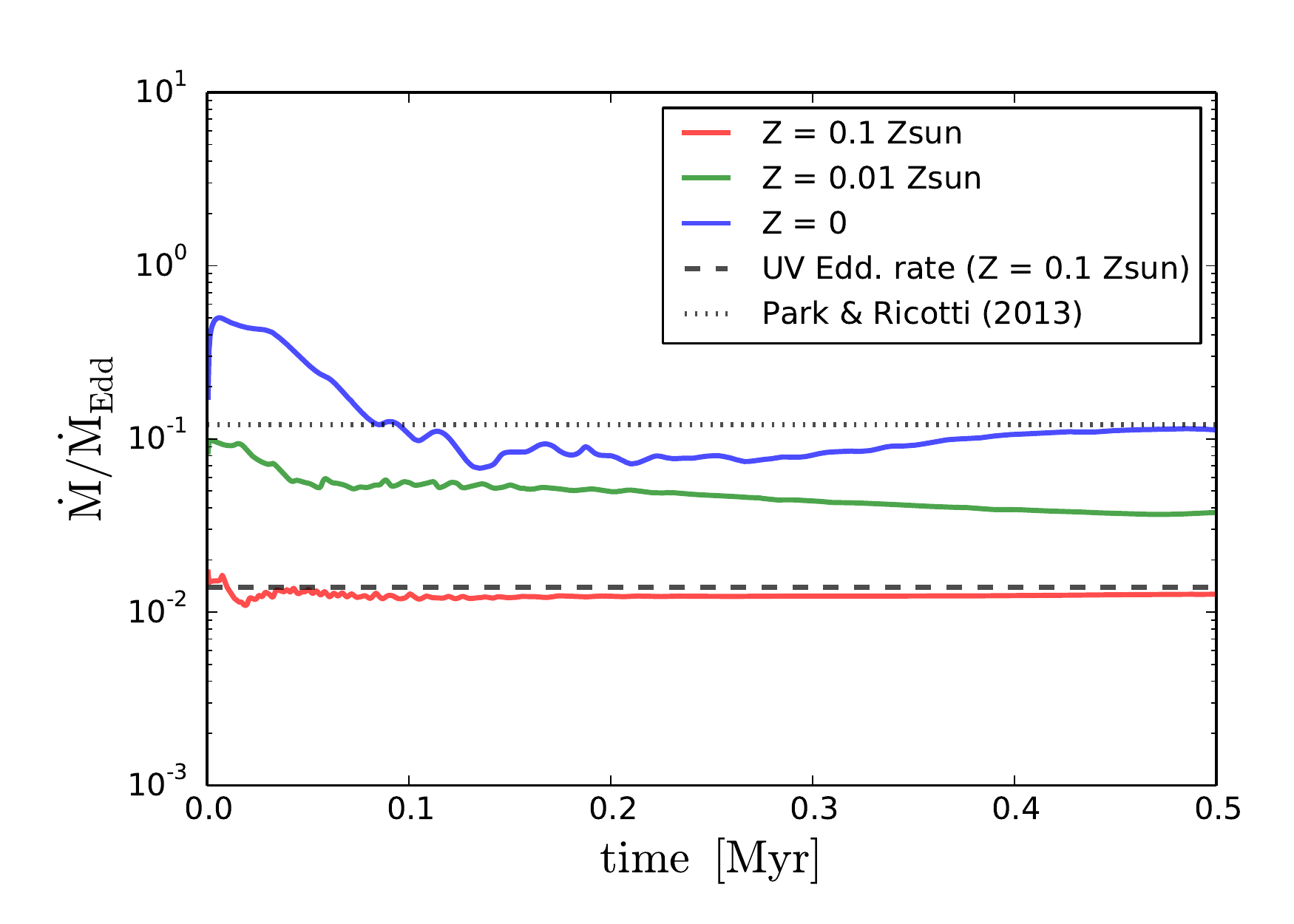}
\end{center}
\caption{
Mass accretion histories onto a moving IMBH with different metallicities, $Z=0$ (PRN4V20, blue), $Z = 10^{-2}~Z_\odot$ (Z2N4V20, green), and $Z = 0.1~Z_\odot$ (Z1N4V20, red). The accretion rate is normalized with the Eddington rate given by Eq. (\ref{eq:medd}) for each case. The dotted line represents the value expected by the 1D analytic model by \protect\cite{Park2013} for the primordial case, where the same BH mass, ambient density, and incident velocity are assumed with $\tinf = 10^4$ K and $\thii = 10^5$ K. The dashed line shows the effective Eddington rate for $Z = 0.1~\zsun$ given by Eq. (\ref{eq:medduv}).
}
\label{fig:mdot_M4N4M2}
\end{figure}

Figure \ref{fig:mdot_M4N4M2} shows the time evolution of mass accretion rates onto the BH for the same cases. For $Z = 0$, the mass accretion rates converge to $\simeq 10~\%$ of the Eddington value, which has been well understood with 1D analytic modeling, as shown in \cite{Park2013}. However, we find that the rates converge to lower values with higher metallicities $Z = 0.01$ and $0.1~\zsun$. This is because the BH gravity is effectively weakened by the outward force of the radiation pressure boosted by the presence of dust grains within the ionized bubble.
Such an effect is incorporated in our Eq. (\ref{eq:medduv}). We see that the rate predicted by Eq. (\ref{eq:medduv}) well matches with the converged rate for $Z = 0.1~\zsun$. 
The lower accretion rate leads to the fainter BH luminosity, which explains the trend in Figure \ref{fig:str_M4N4M2} that the bubble size is smaller with higher metallicity.


\begin{figure}
\begin{center}
\includegraphics[width=8cm]{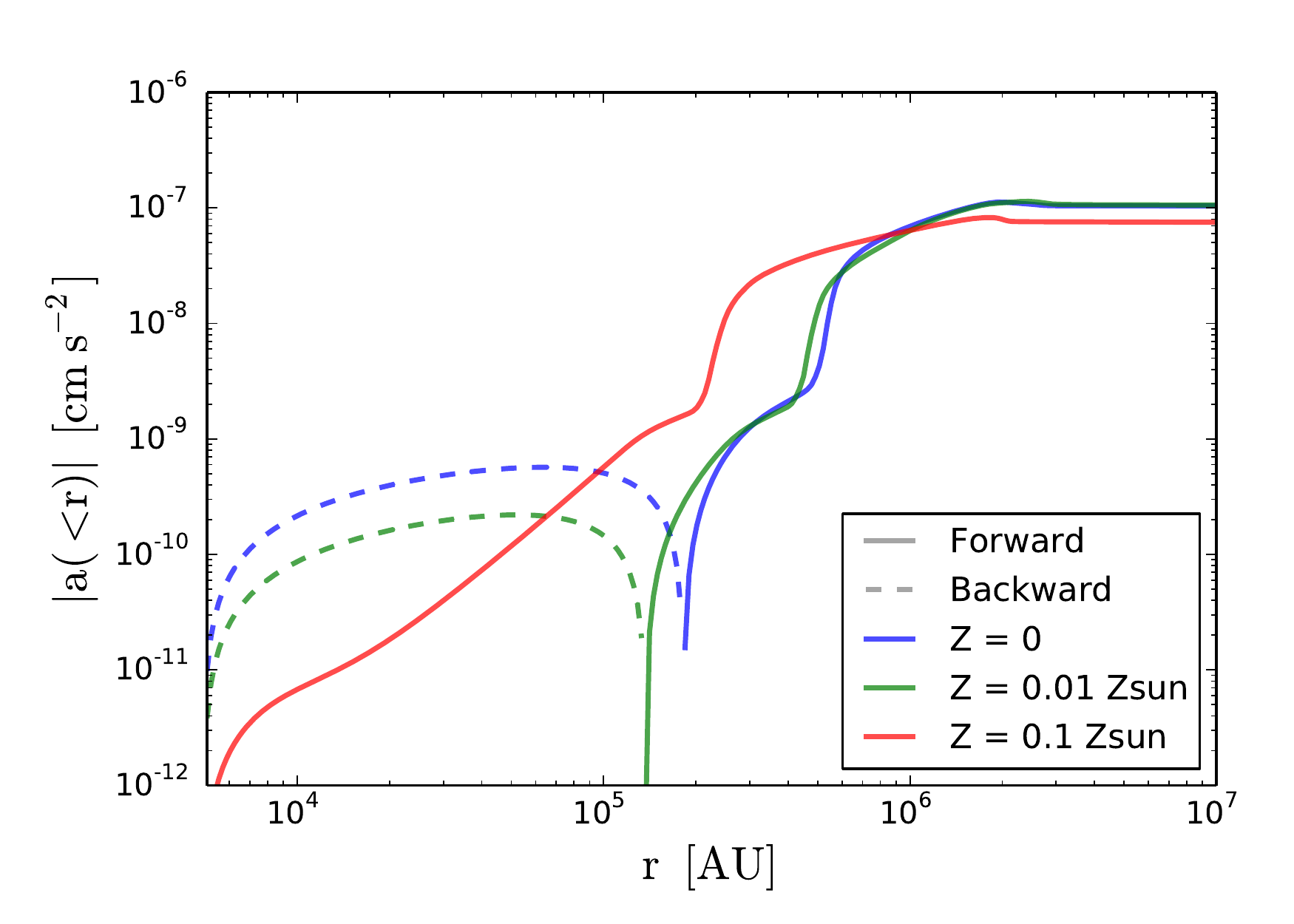}
\end{center}
\caption{
The acceleration of IMBHs caused by the surrounding gas with different metallicities $Z=0$ (PRN4V20, blue), $Z = 10^{-2}~Z_\odot$ (Z2N4V20, green), and $Z = 0.1~Z_\odot$ (Z1N4V20, red), evaluated at the same epoch of $t = 0.4$~Myr. The acceleration is calculated as functions of distance $r$ measured from the BH, for which the gas structure enclosed within the radius $r$ is considered.
The solid and dashed parts of each line represent the forward and backward acceleration, i.e., the IMBH speeds up and down, respectively.
}
\label{fig:ar_M4N4M2}
\end{figure}

Next, we investigate the acceleration of the BH caused by the gravity of the surrounding gas structure. Figure \ref{fig:ar_M4N4M2} presents the BH acceleration evaluated by Eq. (\ref{eq:ay}) for the above runs.
We see that, for $Z = 0$, the gas enclosed within $10^5$ AU contributes to slightly decelerating the BH. That is caused by the downstream wake created in the ionized bubble, similar to the standard BHL case. For $r \gtrsim 10^5$~AU, however, the BH acceleration becomes positive, and its absolute value rapidly increases up to $\ay \sim 10^{-7}~\mr{cm~s^{-2}}$ at $r \sim 10^6$ AU, which corresponds to the position of the dense upstream shell. This result suggests that the BH is gravitationally accelerated forward by the gravity of the dense upstream shell, i.e., the BH speeds up
\footnote{
Although PB17 also suggests the positive BH acceleration, their absolute values are orders of magnitude smaller than our results. Such a large difference is possibly attributable to their erroneous unit conversion (Park 2019, private communication).
}.


Although the flow structure is somewhat different between the cases with $Z = 0$ and $0.01~\zsun$, the resulting acceleration is remarkably similar, particularly for $r \gtrsim 10^5$~AU.
This is because the dense upstream shell structure is common among these cases, and the outer egg-like shell found only for $Z = 0.01~\zsun$ is not massive enough to contribute to the BH acceleration. For $Z = 0.1~\zsun$, the BH acceleration is always positive because the dense upstream shell places much closer to the BH than the other cases. The acceleration still converges to a similar value for $r \gtrsim 10^5$~AU, for which the gravity from the whole part of the dense upstream shell is taken into account.

\begin{figure}
\begin{center}
\includegraphics[width=8cm]{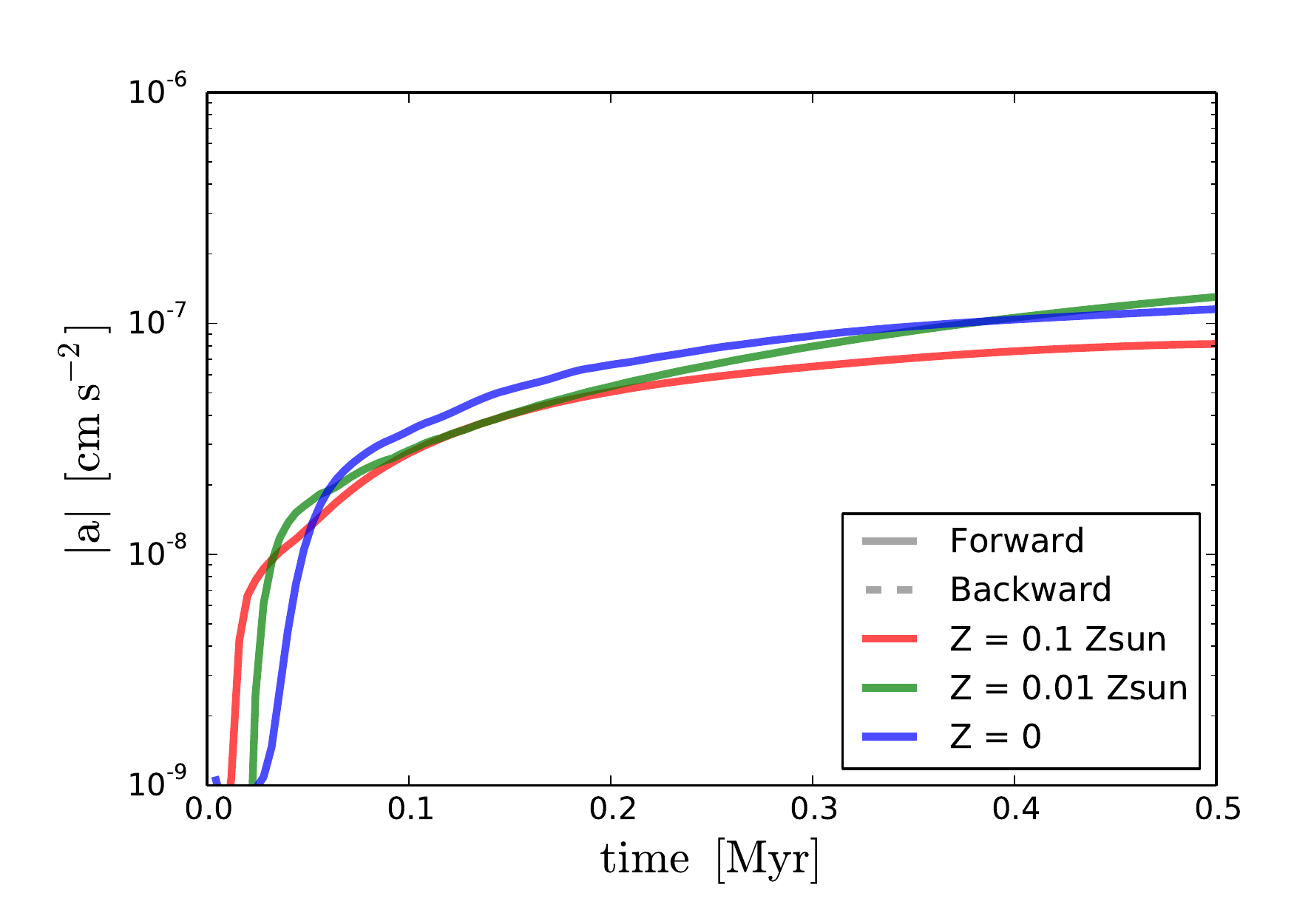}
\end{center}
\caption{
Time evolution of the BH acceleration with different metallicities $Z=0$ (PRN4V20, blue), $Z = 10^{-2}~Z_\odot$ (Z2N4V20, green), and $Z = 0.1~Z_\odot$ (Z1N4V20, red). In this figure, the BH acceleration is evaluated for the whole computational domain, i.e., $r = 10^7$~AU. The plotted lines are all solid, indicating only the forward acceleration is realized for the cases considered.
}
\label{fig:acc_M4N4M2}
\end{figure}

Figure \ref{fig:acc_M4N4M2} presents the time evolution of the BH acceleration caused by the gas contained in the whole computational domain for the above models. The acceleration is always positive over the simulation duration, and it converges to $\ay \sim 10^{-7}~\mr{cm~s^{-2}}$ by the epoch of $t \sim 0.1$ Myr for all cases. For such cases, the timescale over which the BH velocity becomes doubled is estimated as $\tau_\mr{acl} = \vflow/|\ay| \sim 1$ Myr, comparable to the dynamical timescale in galactic disks. It seems difficult for moving IMBHs to fall toward the galactic centers losing their orbital angular momentum.


We here analytically estimate the magnitude of BH acceleration. 
Consider an ionized bubble with a bow-like shocked shell at the upstream front.
As the ram pressure of the pre-shocked gas is converted to the thermal pressure of the post-shocked gas at the shock, the density inside the shell is obtained as $\rho_\mr{shell} \approx (\vflow/c_\mr{shell})^2 \rhoinf$.
Furthermore, according to an analytical consideration based on the mass conservation in the radial and tangential flow inside the shell (Sugimura et al., in prep.), the geometrical thickness of the shell is approximated as $D_\mr{shell} \approx (\rhoinf/\rho_\mr{shell}) \rhii \approx (\vflow/c_\mr{shell})^{-2} \rhii$.
Therefore, the mass of the shell is described as $M_\mr{shell} \approx \pi r^2_\mr{HII} D_\mr{shell} \rho_\mr{shell} \approx \pi \rhoinf r^3_\mr{HII}$.
Finally, we derive the resulting BH acceleration caused by the shell gravity as
\begin{eqnarray}
a \sim \frac{G M_\mr{shell}}{r^2_\mr{HII}} &\sim& \pi G \rhoinf \rhii \nonumber \\
&\sim& 7 \times 10^{-8} \left( \frac{\ninf}{10^4~\mr{cm^{-3}}} \right) \left( \frac{\rhii}{10^6~\mr{AU}} \right) \ ,
\label{eq:a_expect}
\end{eqnarray}
which roughly explains the simulation results.
Furthermore, the dependence of $a \propto \rhii$ in Eq. (\ref{eq:a_expect}) suggests that the smaller bubble size for $Z = 0.1~\zsun$ (Fig.~~\ref{fig:str_M4N4M2}) is the reason for the slightly weaker acceleration than the other cases as shown in Figure \ref{fig:acc_M4N4M2}.

\subsection{Higher Velocity Case} \label{sec:high_vel}

\begin{figure}
\begin{center}
\includegraphics[width=8cm]{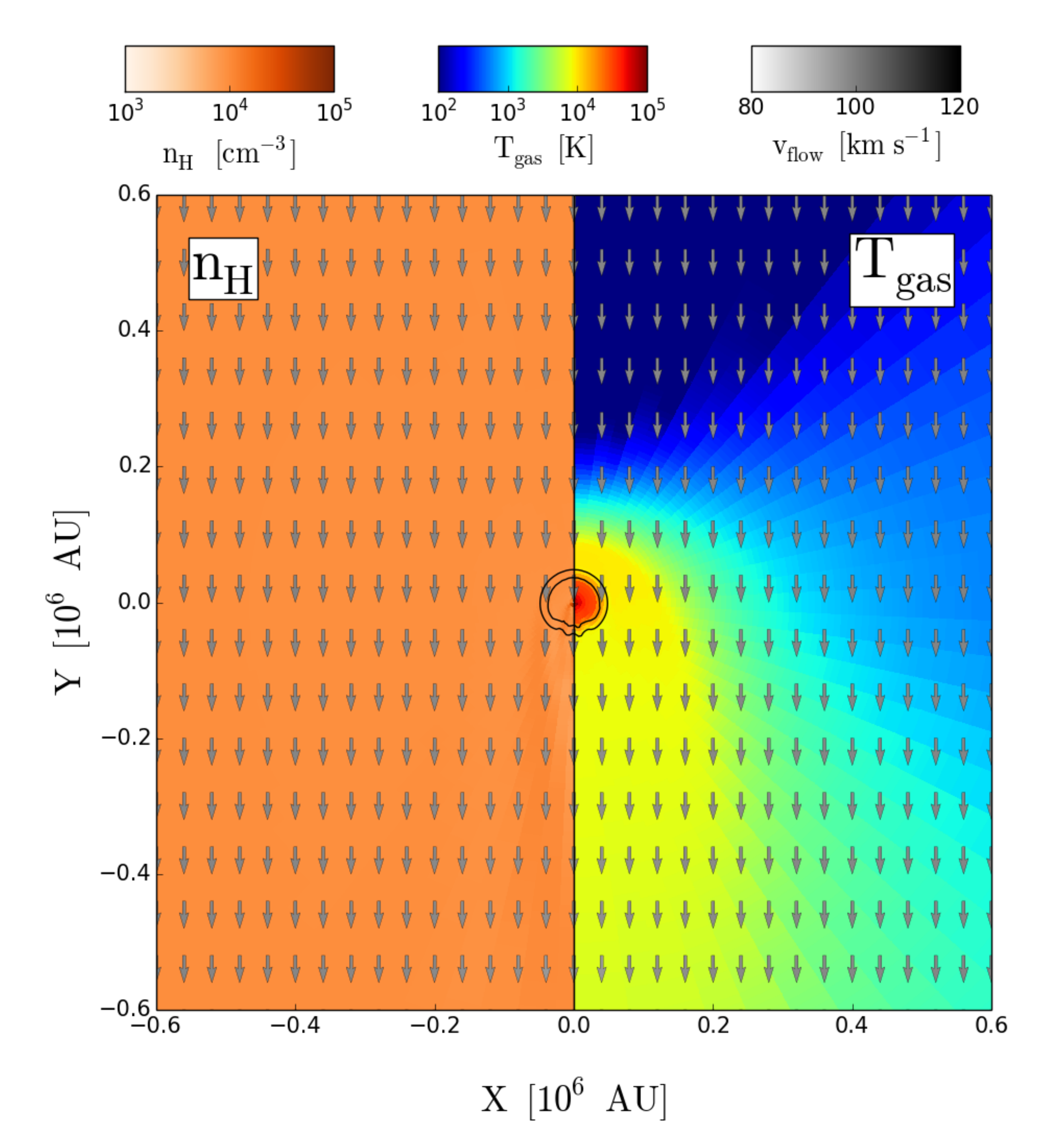}
\end{center}
\caption{
Same as Figure~\ref{fig:str_M4N4M2} but for model Z2N4V100, where the high incident velocity $\vflow = 100~\mr{km~s^{-1}}$ is assumed for the gas with $Z = 0.01~\zsun$. Note that only the color scale of the velocity differs from that in Figure~\ref{fig:str_M4N4M2}. The inner and outer contours indicate the positions where the neutral hydrogen fraction is 0.01 and 0.99. The layer between these contours corresponds to the ionization front.
}
\label{fig:str_M4N4M10}
\end{figure}

As shown above, IMBHs moving at $v_{\rm flow} = 20~\mr{km~s^{-1}}$ should be accelerated owing to the gravitational pull from the dense upstream shell. We here clarify to what extent the acceleration continues, considering the case with the high flow velocity $\vflow = 100~\mr{km~s^{-1}}$. Figure \ref{fig:str_M4N4M10} shows the density and temperature distributions in the quasi-steady state for such a model with $Z = 10^{-2}~Z_\odot$ and $n_\infty = 10^4~{\rm cm}^{-3}$ (model Z2N4V100). In this case, the photoionized bubble, within which the temperature is $\sim 10^5$ K, is much smaller than with $\vflow = 20~\mr{km~s^{-1}}$ (model Z2N4V20, Fig.~\ref{fig:str_M4N4M2}). The small bubble size is due to the low accretion rate (or low BH luminosity), as shown by Eq. (\ref{eq:mb}). Another remarkable difference from model Z2N4V20 is that no dense shell appears ahead of the bubble. The flow is hardly disturbed by the presence of the bubble, as also shown in \cite{Park2013}.
Such flow properties are all expected with the R-type ionization front, a solution realized only for $\vflow \gtrsim \vr = 2 \chii \simeq 60~\mr{km~s^{-1}}$. 
In this case, the flow is too fast for the thermal pressure to create any density jump at the ionization front, consequently prohibiting the formation of the dense upstream shell structure.

\begin{figure}
\begin{center}
\includegraphics[width=8cm]{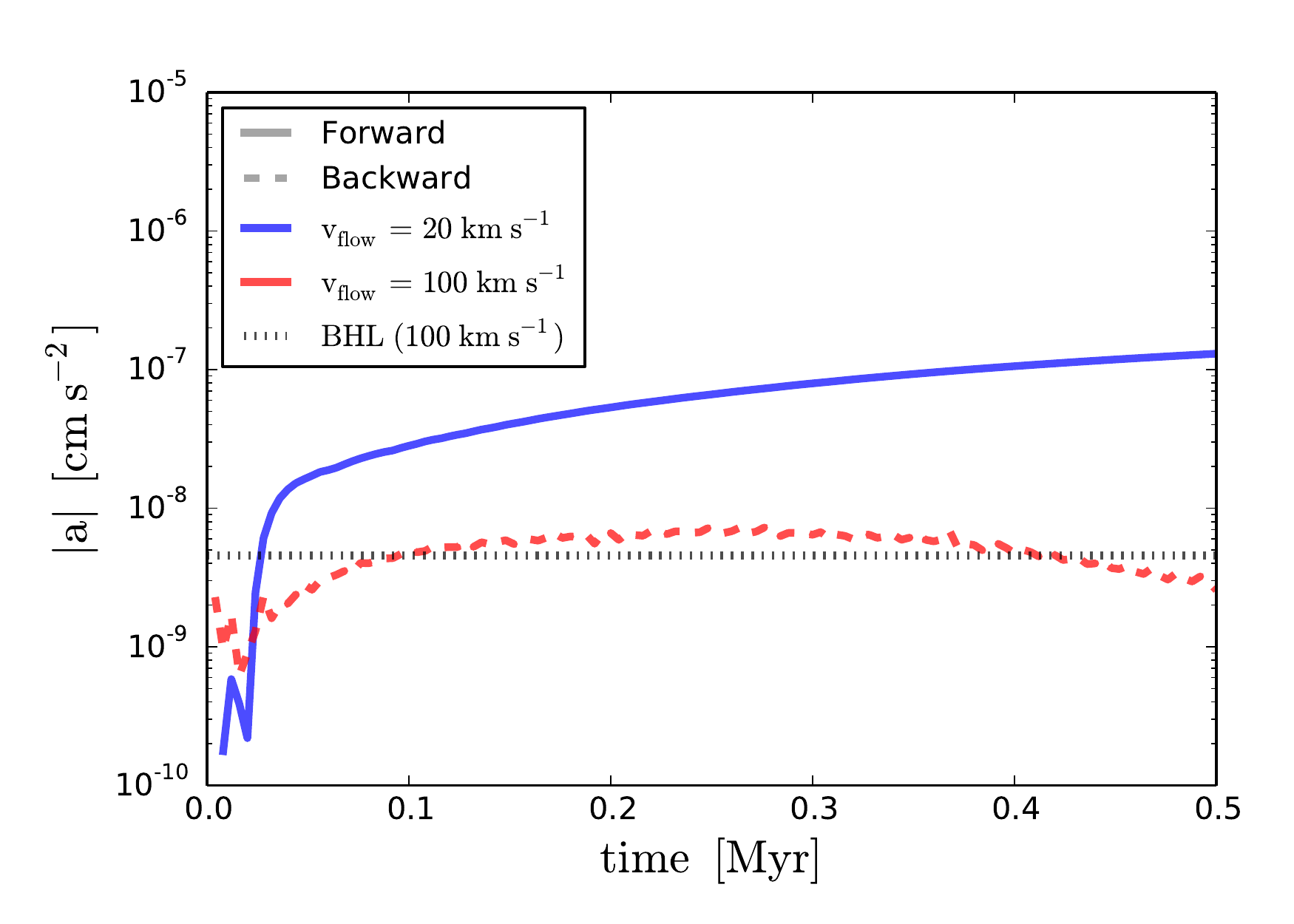}
\end{center}
\caption{
Effects of different incident velocity on the acceleration of the IMBH. The blue and red lines represent the cases with the lower and higher velocities $\vflow = 20~\mr{km~s^{-1}}$ and $100~\mr{km~s^{-1}}$ (models Z2N4V20 and Z2N4V100), respectively. For both cases, the same metallicity $Z= 10^{-2}~\zsun$ is assumed, and the gas contained in the whole computational domain is used to evaluate the acceleration (see Figs.~\ref{fig:ar_M4N4M2} and \ref{fig:acc_M4N4M2}). The dotted line shows the standard BHL dynamical friction, derived with Eq. (\ref{eq:fdfb}) for $\vflow = 100~\mr{km~s^{-1}}$.
}
\label{fig:acc_M4N4M10}
\end{figure}

Figure \ref{fig:acc_M4N4M10} shows the comparison of the resulting BH acceleration between Z2N4V100 and Z2N4V20 models. 
The forward acceleration of moving BHs is no longer expected for the higher velocity case.
In contrast to model Z2N4V20 showing the positive BH acceleration, in model Z2N4V100, the BH rather decelerates at the rate similar to the dynamical friction in the BHL case, which implies that the effects of radiation are almost negligible on the surrounding density structure.
However, the absolute value of the resulting BH acceleration for $\vflow = 100~\mr{km~s^{-1}}$ is much smaller than that for $\vflow = 20~\mr{km~s^{-1}}$, as expected by Eq. (\ref{eq:fdfb}).
Thus, the forward BH acceleration shown in Section \ref{sec:metal} only occurs with $v_{\rm flow} \lesssim \vr \simeq 60~\mr{km~s^{-1}}$, and the acceleration should become less efficient as the velocity approaches $\vr$.

We note here that the negative acceleration predicted in Z2N4V100 model might be modified due to the following reason.
According to the linear analysis by \cite{Newman1967}, the density structure at the R-type ionization front should be unstable.
Actually,  a numerical experiment by Sugimura et al. (in prep.) assuming the constant BH luminosity, but with much higher spatial resolution than models presented here, find the instability of R-type fronts, which may affect the sign and amplitude of the acceleration.
However, in any case, Eq. (\ref{eq:a_expect}) suggests that the BH acceleration would still be negligible because the ionized bubble is small due to inefficient mass accretion onto IMBHs in the case of the high flow velocity. Therefore, we expect our conclusion that the BH acceleration becomes inefficient when $\vflow \gtrsim \vr$ is not affected by the instability.

\begin{figure}
\begin{center}
\includegraphics[width=8cm]{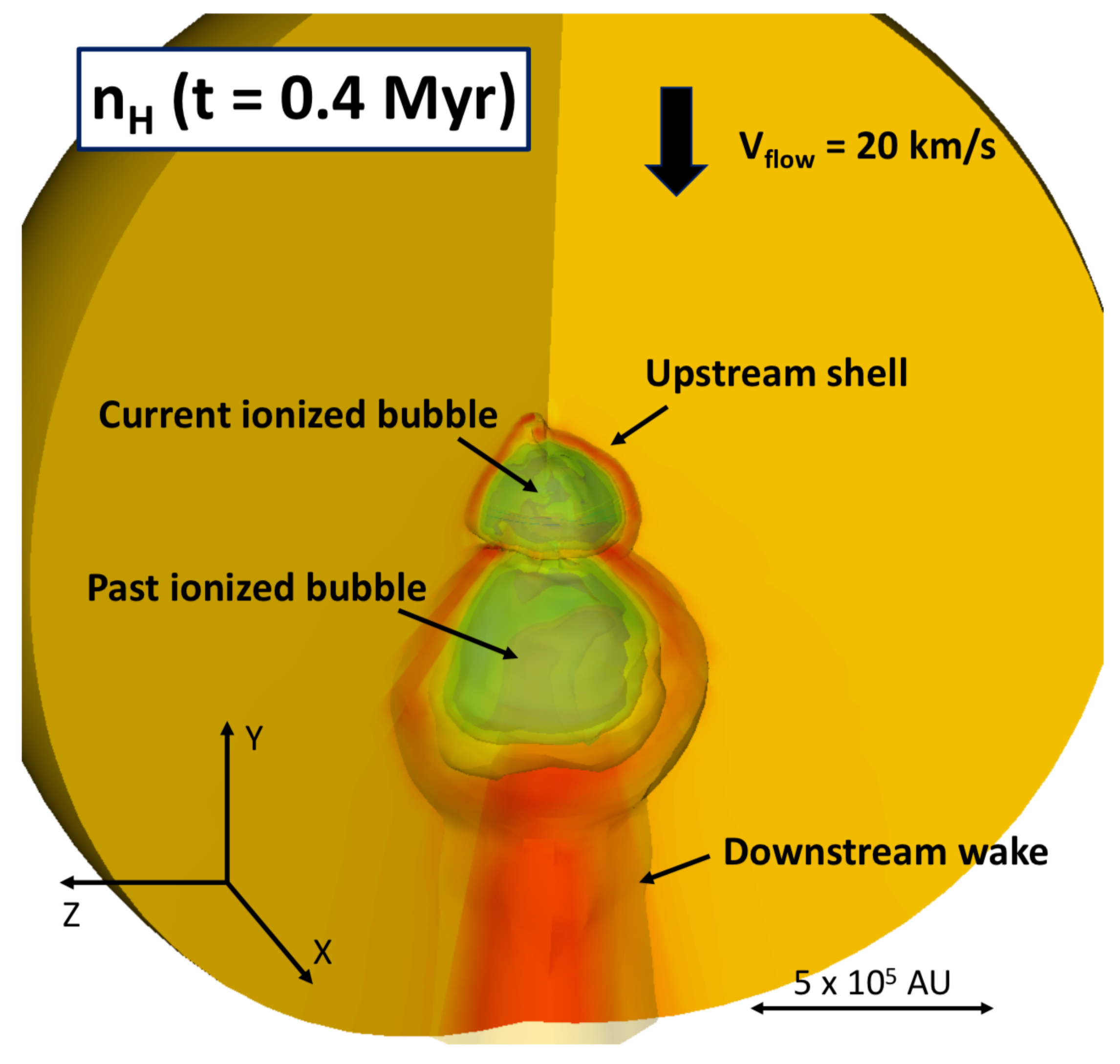}
\end{center}
\caption{
A bird's eye view of the three-dimensional density structure around an IMBH moving through the dense environment with $\ninf = 10^6~\mr{cm^{-3}}$ (model Z2N6V20) for the epoch of $t = 0.4$ Myr.
In this figure, the gas flows downward with $\vflow = 20~\mr{km~s^{-1}}$. The green, orange, and red transparent contours roughly correspond to the isodensity surfaces at $\ninf \sim 10^4,~10^6,$ and $10^7~\mr{cm^{-3}}$, respectively. We see complex structure, consisting of the upstream shell, a current ionized bubble that is excited around the IMBH, and past ionized bubble that is washed away toward the downstream wake (see the text for more details).
}
\label{fig:3d_M4N6M2}
\end{figure}

\begin{figure*}
\begin{center}
\includegraphics[width=17cm]{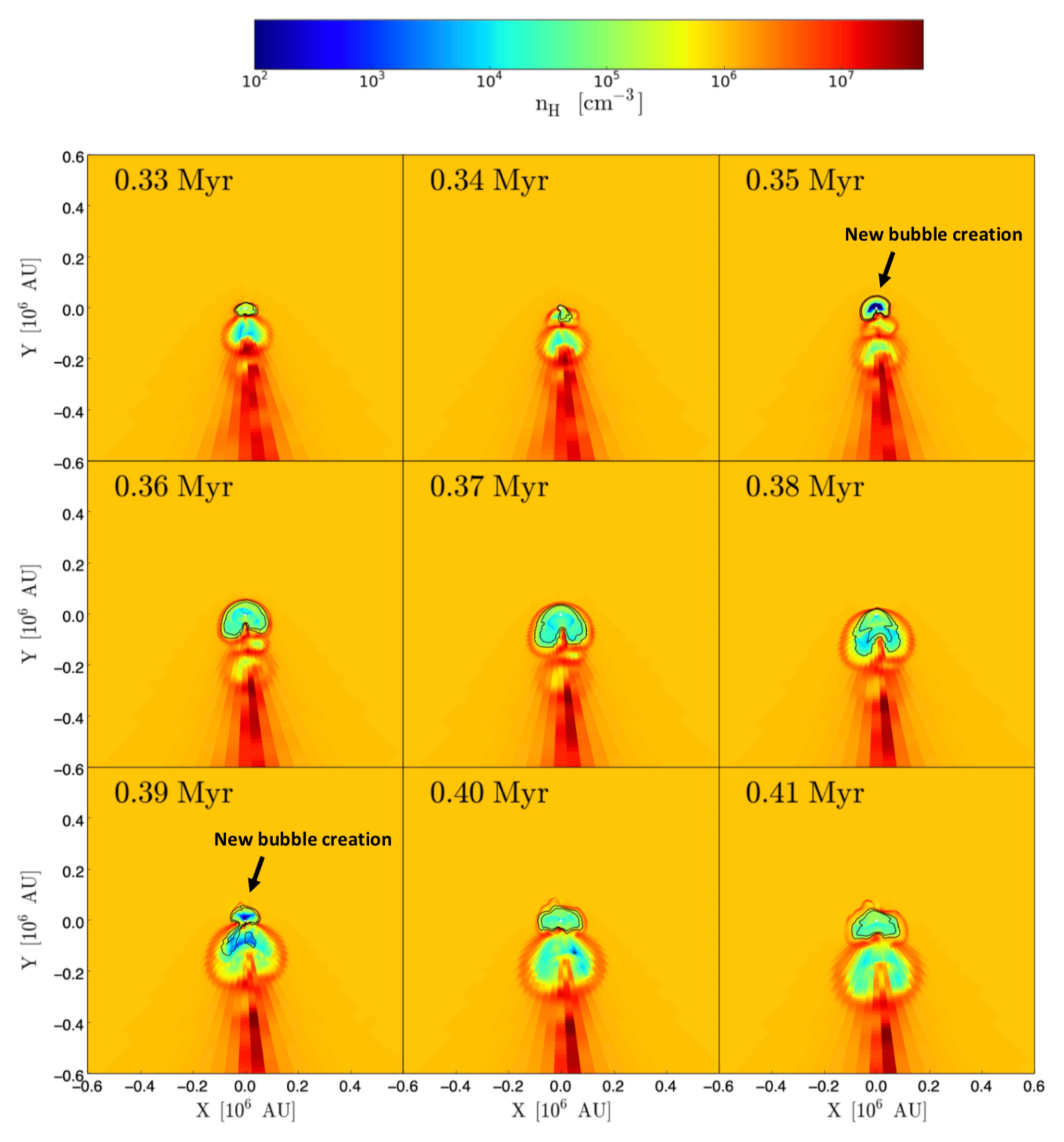}
\end{center}
\caption{
Time variation of the gas density structure in the duration of 0.33 $\leq$ $t$/Myr $\leq$ 0.41 in Z2N6V20 model, where the high ambient density $\ninf = 10^6~\mr{cm^{-3}}$ is assumed. In each panel, the IMBH is located at the center of (X, Y) = (0, 0). The inner and outer contours in each panel indicate the positions where the neutral hydrogen fraction is $10^{-5}$ and 0.99. The layer between these contours corresponds to the ionization front.
}
\label{fig:str_M4N6M2}
\end{figure*}

\begin{figure}
\begin{center}
\includegraphics[width=8cm]{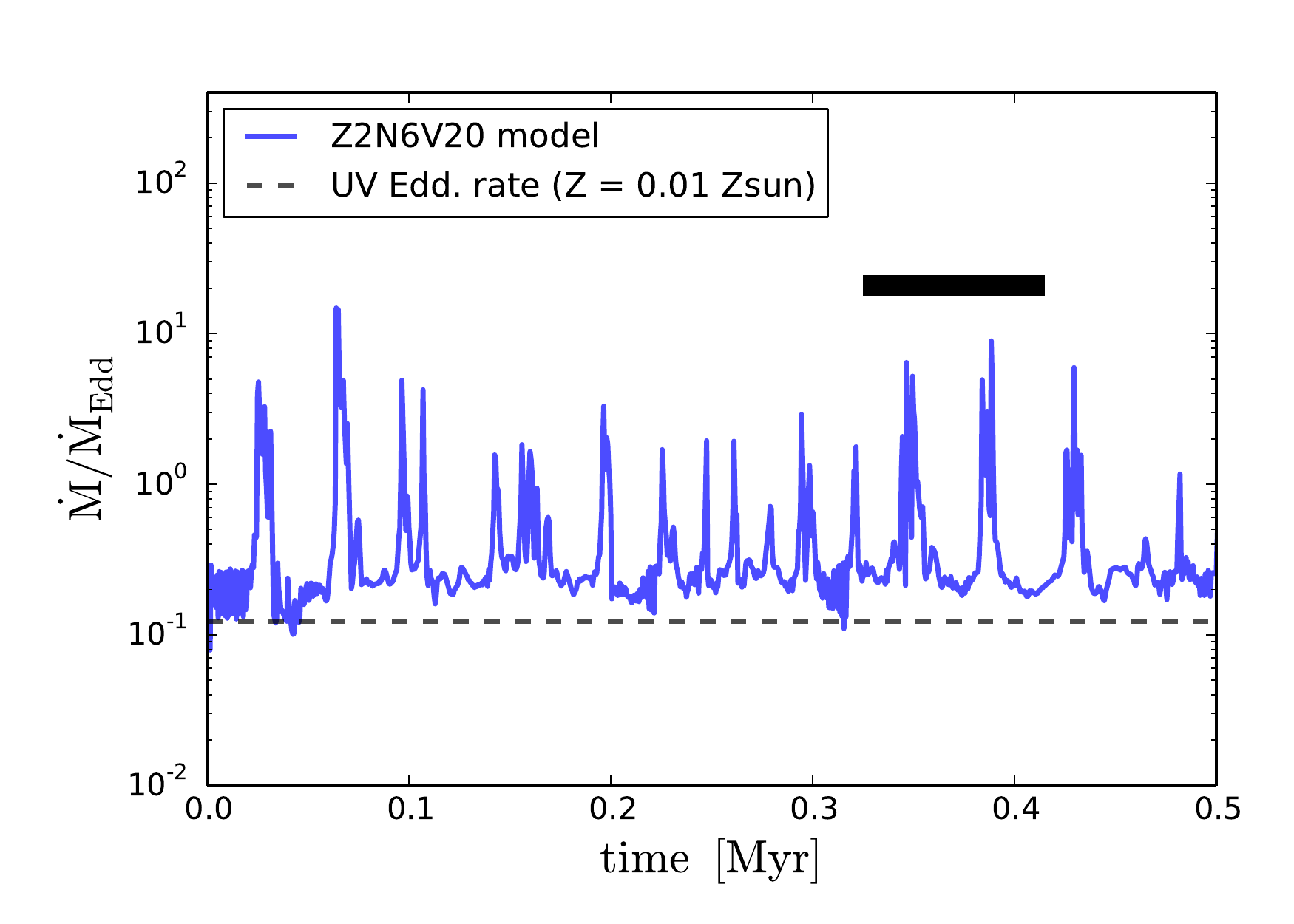}
\end{center}
\caption{
Variable mass accretion histories onto the IMBH observed for Z2N6V20 model. The accretion rate is normalized with the Eddington value as in Figure~\ref{fig:mdot_M4N4M2}. The dashed line shows the effective Eddington rate estimated for the dusty gas with $Z = 0.01~\zsun$ (Eq.~\ref{eq:medduv}). The thick line segment denotes the period of 0.33 $\leq$ $t$/Myr $\leq$ 0.41, for which the temporal variation of the gas density structure is investigated in Figure \ref{fig:str_M4N6M2}.
}
\label{fig:mdot_M4N6M2}
\end{figure}

\begin{figure}
\begin{center}
\includegraphics[width=8cm]{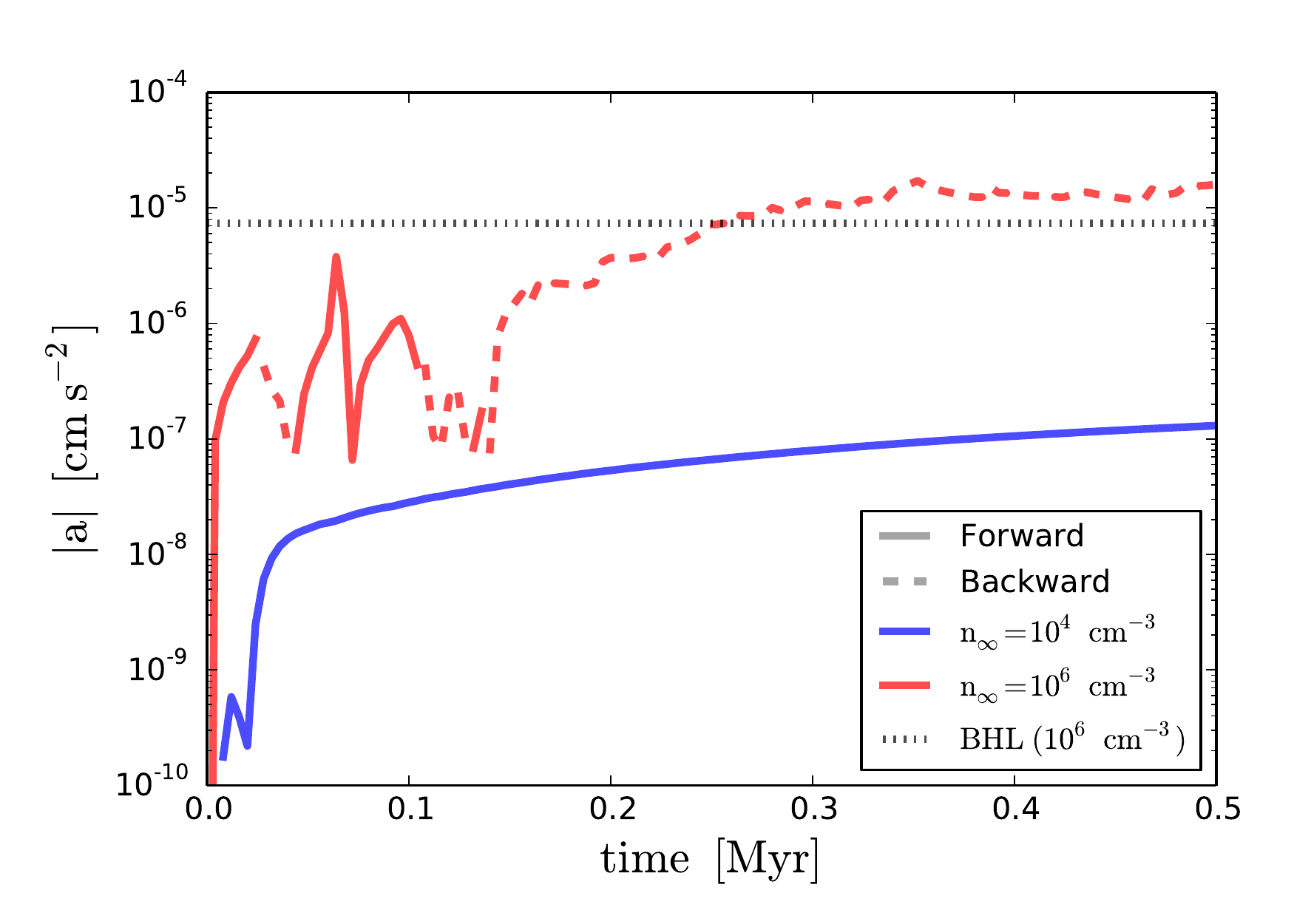}
\end{center}
\caption{
Same as Figure~\ref{fig:acc_M4N4M10} but for effects of the different flow densities. The blue and red lines represent the cases with the lower and higher densities $\ninf = 10^4~\mr{cm^{-3}}$ and $10^6~\mr{cm^{-3}}$ (models Z2N4V20 and Z2N6V100) respectively. We see that only the IMBH moving through the dense environment is decelerated by the frictional force. The dotted line shows the standard BHL dynamical friction, derived with Eq. (\ref{eq:fdfb}) for $\ninf = 10^6~\mr{cm^{-3}}$.
}
\label{fig:acc_M4N6M2}
\end{figure}

\subsection{Dense Environment Case} \label{sec:dense}

We next consider the dense environment where the condition $\mbhf~\ninfs > 1$ is satisfied.
For such a case, the photoionized bubble formation and resulting flow structure around the BH should qualitatively differ from the above cases (see also Sec.~\ref{sec:basic}).
Figure \ref{fig:3d_M4N6M2} shows the three-dimensional density structure obtained for Z2N6V20 model, where the density $\ninf = 10^6~\mr{cm^{-3}}$ is 100 times higher than the other cases. We see that, although some bubble structure associated with the upstream shell appears, the overall structure around the BH is quite different from that found with the rarefied environments (e.g., Figure \ref{fig:str_M4N4M2}). In this model, the ``snowman-like'' double bubble structure is followed by the dense downstream wake.  


We show the time evolution of such structure in Figure \ref{fig:str_M4N6M2}. We find the recurrent bubble formation occurs, i.e., bubbles appear and go downstream one after another. At the epoch of $t$ = 0.35 Myr, a dense shell associated with a bubble hits the central BH, and another bubble newly emerges in response to an induced accretion burst. 
The previous bubble goes downstream, and it deflates being apart from the BH emitting ionizing photons. Such a remnant of the bubble eventually accumulates into the dense downstream wake. The newer bubble also gradually moves downstream, so that the shell in the upstream side approaches the BH in the period of $0.36~\rm{Myr} \lesssim t \lesssim 0.38~\rm{Myr}$.
The shell finally hits the BH at $t$ = 0.39 Myr, and the same evolutionary cycle continues afterwards. Figure \ref{fig:mdot_M4N6M2} shows the mass accretion history onto the BH for the same case. The accretion bursts occur at $t$ = 0.35 and 0.39 Myr, which correspond to the epochs of the new bubble creation presented in Figure \ref{fig:str_M4N6M2}. We see that the accretion rates are enhanced by 10-100 times during the bursts. The figure also shows that such intermittent mass accretion keeps occurring over the whole duration, i.e., the recurrent formation and extinction of the bubble continues while the BH is moving through the dense medium.


Figure \ref{fig:acc_M4N6M2} compares the BH accelerations between the dense ($\ninf = 10^6~\mr{cm^{-3}}$, model Z2N6V20) and rarefied ($\ninf = 10^4~\mr{cm^{-3}}$, model Z2N4V20) environments.
For the dense case, the BH acceleration takes the negative values for $t \gtrsim 0.2$~Myr, when the flow structure reaches the quasi-steady state.
The BH is decelerated in contrast to the rarefied cases described above.
The net backward acceleration is naturally understood because the downstream wake is much larger than the upstream shell. Recall that the acceleration increases with increasing the size of the structure (see Eq. \ref{eq:a_expect}).
The absolute value of the acceleration with $\ninf = 10^6~\mr{cm^{-3}}$ is $|\ay| \sim 10^{-5}~\mr{cm~s^{-2}}$, in good agreement with the dynamical friction in the BHL case with the same density.
This suggests that the radiative feedback no longer affects the orbital evolution of IMBHs in such a dense environment.
The resulting timescale of the BH deceleration is only $\tau_\mr{dec} \sim 0.01$ Myr, much shorter than the dynamical timescale in galactic gas disks. 
We, therefore, conclude that, once an IMBH plunges into a dense environment, it efficiently loses its orbital energy due to strong gaseous dynamical friction.

We note that the unsteady evolution presented above might be caused by the drastic mass accretion onto the BH due to the initial condition far from a hydrodynamical equilibrium. In reality, such rapid accretion can occur when a BH quickly moves from the rarefied to dense environments (see also Sec. \ref{sec:orbital_evol}) and plunges into the massive upstream shell due to the sudden increase of the ram pressure. However, if the ambient density increases more slowly than the timescale over which the BH crosses over the Bondi radius, the mass accretion rate could increase more calmly. To demonstrate such a situation, we additionally performed a numerical experiment, in which the radiative feedback is initially ignored and made effective artificially over the timescale longer than the BH (or the flow) crossing time of the Bondi radius. We found that, in this case, the mass accretion does not occur intermittently but proceeds somewhat steadily, similar to the low-density cases shown in Figure \ref{fig:mdot_M4N4M2}. Interestingly, the flow structure around the BH, however, is much different from one realized in the rarefied environments; the ionized bubble is flown away by the significant ram pressure to be confined within the massive downstream wake.  As a result, the BH experiences almost the same backward acceleration as in Figure \ref{fig:acc_M4N6M2}.  Therefore our conclusion on the gaseous dynamical friction does not change, regardless of whether the intermittent or steady accretion occurs in the dense environments.


\section{DISCUSSIONS} \label{sec:discussion}

\subsection{Can IMBHs Really Accelerate?} \label{sec:can_bh}

In the previous section, we show that IMBHs moving under rarefied environments may speed up until their ionization fronts become R-type.
However, as noted in section \ref{sec:cal_bh_acc}, we do not follow the real-time BH acceleration in our RHD simulations. It is thus still unclear how the properties of dense upstream shell evolve in response to the BH acceleration, and to what extent the BH continues to speed up. We discuss the actual evolution based on our results below.
 

We here consider the dynamical evolution of the BH moving at initial velocity $\vflow < \vr$ with respect to the rarefied ISM satisfying $\mbhf~\ninfs < 1$. In the quasi-steady state realized in our simulation, a dense shell forms ahead of an ionized bubble for such a case (see Section~\ref{sec:metal}). The shell is in a dynamical equilibrium achieved by the balance between the ram pressure upstream $\fram$ and the thermal pressure of the ionized gas $\ftherm$ downstream. In reality, however, the BH gradually accelerates toward the upstream shell. Since the distance between the BH and shell is fixed at the St\"{o}mgren radius, the upstream shell is also accelerated forward in response to the BH acceleration. While the ram pressure $\fram$ increases as the shell accelerates, the thermal pressure $\ftherm$ also increases to satisfy the balance $\fram \sim \ftherm$. This occurs over the sound crossing time of the bubble  $\tau_\mr{sc} = \dsh/\chii \sim 0.1$ Myr, roughly ten times shorter than the timescale of the BH acceleration. Therefore, while the BH speeds up, the upstream shell structure should be approximated to be in the equilibrium states realized finally in our simulations. The acceleration continues until the flow velocity reaches $\vr$.


During the forward acceleration of the BH and shell, gas passing through the ionization front accelerates downstream (see Fig.~\ref{fig:str_M4N4M2}). The gas stays longer in the upstream side than in the downstream side because of the asymmetric velocity structure.
Since the gravitational pull from the BH contributes to increasing the downstream momentum in the upstream side, the flow obtains the net downstream momentum by passing through the bubble around the BH. 
The BH is accelerated forward in return for the backward acceleration of the gas, which ensures the momentum conservation of the whole system. 


The total kinetic energy of the system increases as a result of the acceleration of gas and BH.
Here, the essential drivers of the forward BH acceleration are the dense upstream shell, which is sustained by the thermal energy within the ionized bubble, originating from the energy injection via the BH radiation.
From the viewpoint of energy conservation, this accelerating mechanism is caused by a conversion of gravitational energy of accreting gas to the kinetic energy of the IMBH and its surrounding gas.

\subsection{Orbital Evolution of IMBHs in Merged Galaxies} 
\label{sec:orbital_evol}

Recent numerical simulations demonstrate that highly dense disks should form in the aftermath of the galaxy mergers \citep[e.g.,][]{Mayer2007, Mayer2010, Roskar2015}. Indeed, similar dense gas discs (or CNDs) have been observed in the central parts of ultra-luminous infrared galaxies, which are considered to recently experience merger events \citep[e.g.,][]{Medling2014}.
We here consider the actual orbital evolution of moving IMBHs drifting in such a CND, based on our simulation results.
For an example of the CND, we refer to the disk found in the simulations performed by \cite{Souza-Lima2017}.
They suggest that the CND should have highly clumpy structure created via gravitational instability of the disk. 
Whereas the diffuse gas with $\ninf \lesssim 10^{4}~\mr{cm^{-3}}$ occupies a large volume within the CND, there are small but dense clumps whose typical density is $\ninf \sim 10^{5-7}~\mr{cm^{-3}}$.
The gaseous rotational and turbulent velocities are $\sim 10-100~\mr{km~s^{-1}}$ about 100 pc away from the galactic center.


Consider an IMBH with $\mbh = 10^4~\msun$ moving through the clumpy CND. When the BH is traveling through the diffuse medium, which corresponds to the rarefied environment satisfying $\mbhf~\ninfs < 1$, the BH is accelerated by the gravitational pull from the upstream shell. The forward acceleration continues until the BH velocity relative to the ISM reaches $\vr = 2\chii \sim 60~\mr{km~s^{-1}}$ (Section~\ref{sec:high_vel}).
The BH orbit does not decay during that period. However, while the BH moves around the CND for the several dynamical time $\sim$ 10 Myr, it is expected to encounter a dense clump \citep[][]{Lupi2016}, where the density is high enough to satisfy the condition $\mbhf~\ninfs > 1$. Once this occurs, significant dynamical friction operates on the BH and it shortly loses orbital angular momentum (Section~\ref{sec:dense}).  Our results suggest that spatially resolving the dense clumps is critically important to accurately track the orbital evolution of the BHs drifting in the CND. Even under the radiative feedback, the gaseous dynamical friction should contribute to orbital decay of the BHs, possibly leading to their coalescence in galactic centers.


\subsection{Effects Neglected} \label{sec:multi}

In this section, we discuss the effects that are not taken into account in our RHD simulations but might affect the BH accelerations. 
Namely, we consider the effects of (1) inhomogeneity of surrounding media and (2) anisotropic radiation and mass outflows from the circum-BH accretion disk.

Firstly, we discuss the validity of uniform-gas flow assumed in our RHD simulations.
The CNDs have the disk thickness of $\sim 1-10$ pc, and IMBHs are not necessarily moving along the disk plane, meaning that, in reality, the relative gas flow to IMBHs might be highly variable with time.
For example, an IMBH with a vertical motion to a CND plane experiences an acceleration from the surrounding gas only when the IMBH goes across the disk plane.
Even for an IMBH moving along the CND plane, the gas structure around the IMBH could not be regarded to be uniform.
Actually, for IMBHs with $\mbh = 10^4~\msun$, their Bondi radii and ionized bubbles can be equal to or somewhat larger than the thickness of CND.
In this case, the estimate of the BH acceleration given by Eq. (\ref{eq:a_expect}) should be corrected with the finite volume of the gas disk, leading to the longer BH acceleration timescale than the uniform-gas flow cases. 
Additionally, while we suggested in the previous section that IMBHs should be decelerated by passing through high dense clumps in CNDs, these clumps could also dynamically scatter moving IMBHs lighter than themselves, and prevent them from migrating to galactic centers \citep[][]{Fiacconi2013}.


Secondly, our simulations suppose isotropic radiation and no mass outflow from the circum-BH accretion disk for simplicity.
In reality, because of the absorption of photons along the accretion disk, the resulting radiation field would be anisotropic \citep[][]{Proga2000, Proga2004, Nomura2013, Nomura2016} 
Additionally, mass outflows, such as relativistic jets or line-driven winds, are launched from the vicinity of the central BH toward the polar direction.
In these cases, gas around IMBHs might preferentially distribute along the accretion disk plane, where the effects of radiation and winds are not so significant \citep[][]{Sugimura2017, Takeo2018, Regan2019, Zeilig-Hess2019}.
For accretion disks around moving BHs, since their angular momentum originates from density fluctuation of the flowing gas, its vector is expected to be always perpendicular to the gas flow direction.
Then, the ionized bubble and jets are not able to extend toward the flow direction, and no dense shell structure is expected to form at the upstream side of moving IMBH, as also depicted in Figure 3 of \cite{Gruzinov2019}.
Consequently, moving IMBHs might not accelerate forward, but decelerate similar to the standard BHL case without any radiative or mechanical feedback. 

As seen above, the inhomogeneity in ISM and the existence of anisotropic radiation or jets would provide significant impacts on the orbital evolution of moving IMBHs. Considering these effects is an important subject in our future studies.


\section{SUMMARY AND CONCLUSION} \label{sec:summary}

In this paper, we have investigated how the IMBH with $\mbh = 10^4~\msun$ is dynamically accelerated by the surrounding ISM by performing 3D RHD simulations.
In particular, we have considered IMBHs drifting through the dusty and dense ISM with $Z \leq 0.1~\zsun$ and $\ninf = 10^{4-6}~\mr{cm^{-3}}$, supposing the remnant gas disks left after gas-rich galaxy mergers in the early universe.


First, we have investigated the metallicity dependence of the BH acceleration for the rarefied cases satisfying $\mbhf~\ninfs < 1$. For such cases, regardless of the gas metallicity, our simulations commonly show that a dense shell appears in the upstream side of a photoionized bubble around the BH.
The dense upstream shell gravitationally pulls the moving BH, driving the forward BH acceleration. Such a result is in agreement with previous work by \citetalias{Park2017}, who only considered the primordial case. We show that for the primordial case the magnitude of the BH acceleration is $\sim 10^{-7}~\mr{cm~s^{-2}}$. The resulting timescale of the acceleration is $\sim$ 1 Myr, which is comparable to the dynamical timescale in galactic disks. We have also found that the BH acceleration for $Z = 0.1~\zsun$ is slightly smaller than the primordial case owing to the smaller mass of the upstream shell. The radiation force working on dust grains lowers the accretion rate and BH luminosity, with which the size of the bubble is reduced with finite metallicities. 


In order to clarify to what extent the BH forward acceleration continues, we have further considered the case with the higher flow velocity of $\vflow = 100~\mr{km~s^{-1}}$.
In this case, the ionization front in the upstream side is well approximated with the R-type solution, which is realized for $\vflow \gtrsim \vr = 2 \chii \simeq 60~\mr{km~s^{-1}}$. 
The upstream shell structure no longer appears, so that the forward BH acceleration also disappears. We conclude that only BHs moving with $\vflow < \vr$ accelerate until the velocity approaches $\vr$, above which the acceleration becomes no longer effective.


We have also investigated the BH acceleration under the dense environments satisfying $\mbhf~\ninfs > 1$. In this situation, the flow structure around the BH is totally changed from the other cases. The ionized bubble and the upstream shell are promptly flown downstream by the significant ram pressure of the head wind. The remnants of the bubbles eventually converge to a huge downstream wake, which gravitationally drags the BH backward with the acceleration of $\sim 10^{-5}~\mr{cm~s^{-2}}$. The timescale of the BH deceleration is only $\sim 0.01$ Myr, generally much shorter than the dynamical timescale in galactic gas disks.
Therefore, under the dense environments with $\mbhf~\ninfs > 1$, BHs should quickly lose their orbital energy due to efficient gaseous dynamical friction.


Based on our results, we have discussed the orbital evolution of IMBHs drifting through clumpy CNDs.
When an IMBH with $\mbh = 10^4~\msun$ moves in the medium with $\ninf \lesssim 10^{4}~\mr{cm^{-3}}$, which fills most of the volume of the CND, the BH accelerates forward by the gravity of the upstream shell unless $\vflow > \vr \simeq 60~\mr{km~s^{-1}}$. 
Once the BH encounters a dense clump with $\ninf \gtrsim 10^{6}~\mr{cm^{-3}}$, however, the massive downstream wake appears since the condition of $\mbhf~\ninfs > 1$ is satisfied. In this case, the BH experiences strong dynamical friction from the ambient gas, causing the orbital decay on the much shorter timescale than the dynamical time in the galactic disk.
Therefore, we conclude that although the radiative feedback prevents the orbital decay of BHs in the rarefied environments ($\mbhf~\ninfs < 1$), strong dynamical friction works in dense environments ($\mbhf~\ninfs > 1$), which is likely to be realized after gas-rich galaxy-galaxy mergers. 
The gaseous dynamical friction should contribute to the inward migration of IMBHs even under the radiative feedback, and possibly promote their coalescence in galactic centers.

\section*{Acknowledgements}

The authors would like to thank Naoki Yoshida, Pratika Dayal, and Raffaella Schneider for fruitful discussions, and Riouhei Nakatani for his contribution to developing the numerical code. The numerical simulations were performed on the Cray XC50 at the Center for Computational Astrophysics (CfCA) of the National Astronomical Observatory of Japan. This work is financially supported by the Grants-in-Aid for Basic Research by the Ministry of Education, Science and Culture of Japan (17H06360: D.T., 16H05996, 19H01934: T.H.). K.S. appreciates the support by the JSPS Overseas Research Fellowship. R.K. acknowledges financial support via the Emmy Noether Research Group on Accretion Flows and Feedback in Realistic Models of Massive Star Formation funded by the German Research Foundation (DFG) under grant no. KU 2849/3-1 and KU 2849/3-2.

\bibliographystyle{mnras}
\bibliography{ref}

\begin{thebibliography}{}
\makeatletter
\relax
\def\mn@urlcharsother{\let\do\@makeother \do\$\do\&\do\#\do\^\do\_\do\%\do\~}
\def\mn@doi{\begingroup\mn@urlcharsother \@ifnextchar [ {\mn@doi@}
  {\mn@doi@[]}}
\def\mn@doi@[#1]#2{\def\@tempa{#1}\ifx\@tempa\@empty \href
  {http://dx.doi.org/#2} {doi:#2}\else \href {http://dx.doi.org/#2} {#1}\fi
  \endgroup}
\def\mn@eprint#1#2{\mn@eprint@#1:#2::\@nil}
\def\mn@eprint@arXiv#1{\href {http://arxiv.org/abs/#1} {{\tt arXiv:#1}}}
\def\mn@eprint@dblp#1{\href {http://dblp.uni-trier.de/rec/bibtex/#1.xml}
  {dblp:#1}}
\def\mn@eprint@#1:#2:#3:#4\@nil{\def\@tempa {#1}\def\@tempb {#2}\def\@tempc
  {#3}\ifx \@tempc \@empty \let \@tempc \@tempb \let \@tempb \@tempa \fi \ifx
  \@tempb \@empty \def\@tempb {arXiv}\fi \@ifundefined
  {mn@eprint@\@tempb}{\@tempb:\@tempc}{\expandafter \expandafter \csname
  mn@eprint@\@tempb\endcsname \expandafter{\@tempc}}}

\bibitem[\protect\citeauthoryear{{Amaro-Seoane} et~al.,}{{Amaro-Seoane}
  et~al.}{2017}]{LISA2017}
{Amaro-Seoane} P.,  et~al., 2017, arXiv e-prints, \href
  {https://ui.adsabs.harvard.edu/abs/2017arXiv170200786A} {p. arXiv:1702.00786}

\bibitem[\protect\citeauthoryear{{Ba{\~n}ados} et~al.,}{{Ba{\~n}ados}
  et~al.}{2018}]{Banados2018}
{Ba{\~n}ados} E.,  et~al., 2018, \mn@doi [\nat] {10.1038/nature25180}, \href
  {https://ui.adsabs.harvard.edu/abs/2018Natur.553..473B} {553, 473}

\bibitem[\protect\citeauthoryear{{Davies}, {Mark}  \& {Sternberg}}{{Davies}
  et~al.}{2012}]{Davies2012}
{Davies} R.,  {Mark} D.,   {Sternberg} A.,  2012, \mn@doi [\aap]
  {10.1051/0004-6361/201117647}, \href
  {https://ui.adsabs.harvard.edu/abs/2012A&A...537A.133D} {537, A133}

\bibitem[\protect\citeauthoryear{{Di Matteo}, {Croft}, {Feng}, {Waters}  \&
  {Wilkins}}{{Di Matteo} et~al.}{2017}]{Di_Matteo2017}
{Di Matteo} T.,  {Croft} R. A.~C.,  {Feng} Y.,  {Waters} D.,   {Wilkins} S.,
  2017, \mn@doi [\mnras] {10.1093/mnras/stx319}, \href
  {https://ui.adsabs.harvard.edu/abs/2017MNRAS.467.4243D} {467, 4243}

\bibitem[\protect\citeauthoryear{{Escala}, {Larson}, {Coppi}  \&
  {Mardones}}{{Escala} et~al.}{2005}]{Escala2005}
{Escala} A.,  {Larson} R.~B.,  {Coppi} P.~S.,   {Mardones} D.,  2005, \mn@doi
  [\apj] {10.1086/431747}, \href
  {https://ui.adsabs.harvard.edu/abs/2005ApJ...630..152E} {630, 152}

\bibitem[\protect\citeauthoryear{{Fan} et~al.,}{{Fan} et~al.}{2001}]{Fan2001}
{Fan} X.,  et~al., 2001, \mn@doi [\aj] {10.1086/324111}, \href
  {https://ui.adsabs.harvard.edu/abs/2001AJ....122.2833F} {122, 2833}

\bibitem[\protect\citeauthoryear{{Fiacconi}, {Mayer}, {Ro{\v{s}}kar}  \&
  {Colpi}}{{Fiacconi} et~al.}{2013}]{Fiacconi2013}
{Fiacconi} D.,  {Mayer} L.,  {Ro{\v{s}}kar} R.,   {Colpi} M.,  2013, \mn@doi
  [\apjl] {10.1088/2041-8205/777/1/L14}, \href
  {https://ui.adsabs.harvard.edu/abs/2013ApJ...777L..14F} {777, L14}

\bibitem[\protect\citeauthoryear{{Gruzinov}, {Levin}  \& {Matzner}}{{Gruzinov}
  et~al.}{2019}]{Gruzinov2019}
{Gruzinov} A.,  {Levin} Y.,   {Matzner} C.~D.,  2019, arXiv e-prints, \href
  {https://ui.adsabs.harvard.edu/abs/2019arXiv190601186G} {p. arXiv:1906.01186}

\bibitem[\protect\citeauthoryear{{Hopkins}, {Hernquist}, {Cox}  \&
  {Kere{\v{s}}}}{{Hopkins} et~al.}{2008}]{Hopkins2008}
{Hopkins} P.~F.,  {Hernquist} L.,  {Cox} T.~J.,   {Kere{\v{s}}} D.,  2008,
  \mn@doi [\apjs] {10.1086/524362}, \href
  {https://ui.adsabs.harvard.edu/abs/2008ApJS..175..356H} {175, 356}

\bibitem[\protect\citeauthoryear{{Hosokawa}, {Hirano}, {Kuiper}, {Yorke},
  {Omukai}  \& {Yoshida}}{{Hosokawa} et~al.}{2016}]{Hosokawa2016}
{Hosokawa} T.,  {Hirano} S.,  {Kuiper} R.,  {Yorke} H.~W.,  {Omukai} K.,
  {Yoshida} N.,  2016, \mn@doi [\apj] {10.3847/0004-637X/824/2/119}, \href
  {https://ui.adsabs.harvard.edu/abs/2016ApJ...824..119H} {824, 119}

\bibitem[\protect\citeauthoryear{{Huang}, {Feng}  \& {Di Matteo}}{{Huang}
  et~al.}{2019}]{Huang2019}
{Huang} K.-W.,  {Feng} Y.,   {Di Matteo} T.,  2019, arXiv e-prints, \href
  {https://ui.adsabs.harvard.edu/abs/2019arXiv190600242H} {p. arXiv:1906.00242}

\bibitem[\protect\citeauthoryear{{Inayoshi}, {Haiman}  \&
  {Ostriker}}{{Inayoshi} et~al.}{2016}]{Inayoshi2016}
{Inayoshi} K.,  {Haiman} Z.,   {Ostriker} J.~P.,  2016, \mn@doi [\mnras]
  {10.1093/mnras/stw836}, \href
  {https://ui.adsabs.harvard.edu/abs/2016MNRAS.459.3738I} {459, 3738}

\bibitem[\protect\citeauthoryear{{Inayoshi}, {Visbal}  \& {Haiman}}{{Inayoshi}
  et~al.}{2019}]{Inayoshi2019}
{Inayoshi} K.,  {Visbal} E.,   {Haiman} Z.,  2019, arXiv e-prints, \href
  {https://ui.adsabs.harvard.edu/abs/2019arXiv191105791I} {p. arXiv:1911.05791}

\bibitem[\protect\citeauthoryear{{Izumi} et~al.,}{{Izumi}
  et~al.}{2013}]{Izumi2013}
{Izumi} T.,  et~al., 2013, \mn@doi [\pasj] {10.1093/pasj/65.5.100}, \href
  {https://ui.adsabs.harvard.edu/abs/2013PASJ...65..100I} {65, 100}

\bibitem[\protect\citeauthoryear{{K{\"o}lligan} \& {Kuiper}}{{K{\"o}lligan} \&
  {Kuiper}}{2018}]{Kolligan2018}
{K{\"o}lligan} A.,  {Kuiper} R.,  2018, \mn@doi [\aap]
  {10.1051/0004-6361/201833686}, \href
  {https://ui.adsabs.harvard.edu/abs/2018A&A...620A.182K} {620, A182}

\bibitem[\protect\citeauthoryear{{Kuiper} \& {Hosokawa}}{{Kuiper} \&
  {Hosokawa}}{2018}]{Kuiper2018}
{Kuiper} R.,  {Hosokawa} T.,  2018, \mn@doi [\aap]
  {10.1051/0004-6361/201832638}, \href
  {https://ui.adsabs.harvard.edu/abs/2018A&A...616A.101K} {616, A101}

\bibitem[\protect\citeauthoryear{{Kuiper}, {Klahr}, {Beuther}  \&
  {Henning}}{{Kuiper} et~al.}{2010}]{Kuiper2010}
{Kuiper} R.,  {Klahr} H.,  {Beuther} H.,   {Henning} T.,  2010, \mn@doi [\apj]
  {10.1088/0004-637X/722/2/1556}, \href
  {https://ui.adsabs.harvard.edu/abs/2010ApJ...722.1556K} {722, 1556}

\bibitem[\protect\citeauthoryear{{Kuiper}, {Klahr}, {Beuther}  \&
  {Henning}}{{Kuiper} et~al.}{2011}]{Kuiper2011}
{Kuiper} R.,  {Klahr} H.,  {Beuther} H.,   {Henning} T.,  2011, \mn@doi [\apj]
  {10.1088/0004-637X/732/1/20}, \href
  {https://ui.adsabs.harvard.edu/abs/2011ApJ...732...20K} {732, 20}

\bibitem[\protect\citeauthoryear{{Kuiper}, {Klahr}, {Beuther}  \&
  {Henning}}{{Kuiper} et~al.}{2012}]{Kuiper2012}
{Kuiper} R.,  {Klahr} H.,  {Beuther} H.,   {Henning} T.,  2012, \mn@doi [\aap]
  {10.1051/0004-6361/201117808}, \href
  {https://ui.adsabs.harvard.edu/abs/2012A&A...537A.122K} {537, A122}

\bibitem[\protect\citeauthoryear{{Lupi}, {Haardt}, {Dotti}, {Fiacconi}, {Mayer}
   \& {Madau}}{{Lupi} et~al.}{2016}]{Lupi2016}
{Lupi} A.,  {Haardt} F.,  {Dotti} M.,  {Fiacconi} D.,  {Mayer} L.,   {Madau}
  P.,  2016, \mn@doi [\mnras] {10.1093/mnras/stv2877}, \href
  {https://ui.adsabs.harvard.edu/abs/2016MNRAS.456.2993L} {456, 2993}

\bibitem[\protect\citeauthoryear{{Matsuoka} et~al.,}{{Matsuoka}
  et~al.}{2019}]{Matsuoka2019}
{Matsuoka} Y.,  et~al., 2019, \mn@doi [\apjl] {10.3847/2041-8213/ab0216}, \href
  {https://ui.adsabs.harvard.edu/abs/2019ApJ...872L...2M} {872, L2}

\bibitem[\protect\citeauthoryear{{Mayer}, {Kazantzidis}, {Madau}, {Colpi},
  {Quinn}  \& {Wadsley}}{{Mayer} et~al.}{2007}]{Mayer2007}
{Mayer} L.,  {Kazantzidis} S.,  {Madau} P.,  {Colpi} M.,  {Quinn} T.,
  {Wadsley} J.,  2007, \mn@doi [Science] {10.1126/science.1141858}, \href
  {https://ui.adsabs.harvard.edu/abs/2007Sci...316.1874M} {316, 1874}

\bibitem[\protect\citeauthoryear{{Mayer}, {Kazantzidis}, {Escala}  \&
  {Callegari}}{{Mayer} et~al.}{2010}]{Mayer2010}
{Mayer} L.,  {Kazantzidis} S.,  {Escala} A.,   {Callegari} S.,  2010, \mn@doi
  [\nat] {10.1038/nature09294}, \href
  {https://ui.adsabs.harvard.edu/abs/2010Natur.466.1082M} {466, 1082}

\bibitem[\protect\citeauthoryear{{Medling} et~al.,}{{Medling}
  et~al.}{2014}]{Medling2014}
{Medling} A.~M.,  et~al., 2014, \mn@doi [\apj] {10.1088/0004-637X/784/1/70},
  \href {https://ui.adsabs.harvard.edu/abs/2014ApJ...784...70M} {784, 70}

\bibitem[\protect\citeauthoryear{{Mignone}, {Bodo}, {Massaglia}, {Matsakos},
  {Tesileanu}, {Zanni}  \& {Ferrari}}{{Mignone} et~al.}{2007}]{Mignone2007}
{Mignone} A.,  {Bodo} G.,  {Massaglia} S.,  {Matsakos} T.,  {Tesileanu} O.,
  {Zanni} C.,   {Ferrari} A.,  2007, \mn@doi [\apjs] {10.1086/513316}, \href
  {https://ui.adsabs.harvard.edu/abs/2007ApJS..170..228M} {170, 228}

\bibitem[\protect\citeauthoryear{{Mortlock} et~al.,}{{Mortlock}
  et~al.}{2011}]{Mortlock2011}
{Mortlock} D.~J.,  et~al., 2011, \mn@doi [\nat] {10.1038/nature10159}, \href
  {https://ui.adsabs.harvard.edu/abs/2011Natur.474..616M} {474, 616}

\bibitem[\protect\citeauthoryear{{Nakatani}, {Hosokawa}, {Yoshida}, {Nomura}
  \& {Kuiper}}{{Nakatani} et~al.}{2018a}]{Nakatani2018a}
{Nakatani} R.,  {Hosokawa} T.,  {Yoshida} N.,  {Nomura} H.,   {Kuiper} R.,
  2018a, \mn@doi [\apj] {10.3847/1538-4357/aab70b}, \href
  {https://ui.adsabs.harvard.edu/abs/2018ApJ...857...57N} {857, 57}

\bibitem[\protect\citeauthoryear{{Nakatani}, {Hosokawa}, {Yoshida}, {Nomura}
  \& {Kuiper}}{{Nakatani} et~al.}{2018b}]{Nakatani2018b}
{Nakatani} R.,  {Hosokawa} T.,  {Yoshida} N.,  {Nomura} H.,   {Kuiper} R.,
  2018b, \mn@doi [\apj] {10.3847/1538-4357/aad9fd}, \href
  {https://ui.adsabs.harvard.edu/abs/2018ApJ...865...75N} {865, 75}

\bibitem[\protect\citeauthoryear{{Newman} \& {Axford}}{{Newman} \&
  {Axford}}{1967}]{Newman1967}
{Newman} R.~C.,  {Axford} W.~I.,  1967, \mn@doi [\apj] {10.1086/149286}, \href
  {https://ui.adsabs.harvard.edu/abs/1967ApJ...149..571N} {149, 571}

\bibitem[\protect\citeauthoryear{{Nomura}, {Ohsuga}, {Wada}, {Susa}  \&
  {Misawa}}{{Nomura} et~al.}{2013}]{Nomura2013}
{Nomura} M.,  {Ohsuga} K.,  {Wada} K.,  {Susa} H.,   {Misawa} T.,  2013,
  \mn@doi [\pasj] {10.1093/pasj/65.2.40}, \href
  {https://ui.adsabs.harvard.edu/abs/2013PASJ...65...40N} {65, 40}

\bibitem[\protect\citeauthoryear{{Nomura}, {Ohsuga}, {Takahashi}, {Wada}  \&
  {Yoshida}}{{Nomura} et~al.}{2016}]{Nomura2016}
{Nomura} M.,  {Ohsuga} K.,  {Takahashi} H.~R.,  {Wada} K.,   {Yoshida} T.,
  2016, \mn@doi [\pasj] {10.1093/pasj/psv124}, \href
  {https://ui.adsabs.harvard.edu/abs/2016PASJ...68...16N} {68, 16}

\bibitem[\protect\citeauthoryear{{Osterbrock}}{{Osterbrock}}{1989}]{Osterbrock1989}
{Osterbrock} D.~E.,  1989, {Astrophysics of gaseous nebulae and active galactic
  nuclei}

\bibitem[\protect\citeauthoryear{{Ostriker}}{{Ostriker}}{1999}]{Ostriker1999}
{Ostriker} E.~C.,  1999, \mn@doi [\apj] {10.1086/306858}, \href
  {https://ui.adsabs.harvard.edu/abs/1999ApJ...513..252O} {513, 252}

\bibitem[\protect\citeauthoryear{{Park} \& {Bogdanovi{\'c}}}{{Park} \&
  {Bogdanovi{\'c}}}{2017}]{Park2017}
{Park} K.,  {Bogdanovi{\'c}} T.,  2017, \mn@doi [\apj]
  {10.3847/1538-4357/aa65ce}, \href
  {https://ui.adsabs.harvard.edu/abs/2017ApJ...838..103P} {838, 103}

\bibitem[\protect\citeauthoryear{{Park} \& {Ricotti}}{{Park} \&
  {Ricotti}}{2013}]{Park2013}
{Park} K.,  {Ricotti} M.,  2013, \mn@doi [\apj] {10.1088/0004-637X/767/2/163},
  \href {https://ui.adsabs.harvard.edu/abs/2013ApJ...767..163P} {767, 163}

\bibitem[\protect\citeauthoryear{{Proga} \& {Kallman}}{{Proga} \&
  {Kallman}}{2004}]{Proga2004}
{Proga} D.,  {Kallman} T.~R.,  2004, \mn@doi [\apj] {10.1086/425117}, \href
  {https://ui.adsabs.harvard.edu/abs/2004ApJ...616..688P} {616, 688}

\bibitem[\protect\citeauthoryear{{Proga}, {Stone}  \& {Kallman}}{{Proga}
  et~al.}{2000}]{Proga2000}
{Proga} D.,  {Stone} J.~M.,   {Kallman} T.~R.,  2000, \mn@doi [\apj]
  {10.1086/317154}, \href
  {https://ui.adsabs.harvard.edu/abs/2000ApJ...543..686P} {543, 686}

\bibitem[\protect\citeauthoryear{{Regan}, {Downes}, {Volonteri}, {Beckmann},
  {Lupi}, {Trebitsch}  \& {Dubois}}{{Regan} et~al.}{2019}]{Regan2019}
{Regan} J.~A.,  {Downes} T.~P.,  {Volonteri} M.,  {Beckmann} R.,  {Lupi} A.,
  {Trebitsch} M.,   {Dubois} Y.,  2019, \mn@doi [\mnras]
  {10.1093/mnras/stz1045}, \href
  {https://ui.adsabs.harvard.edu/abs/2019MNRAS.486.3892R} {486, 3892}

\bibitem[\protect\citeauthoryear{{Ro{\v{s}}kar}, {Fiacconi}, {Mayer},
  {Kazantzidis}, {Quinn}  \& {Wadsley}}{{Ro{\v{s}}kar}
  et~al.}{2015}]{Roskar2015}
{Ro{\v{s}}kar} R.,  {Fiacconi} D.,  {Mayer} L.,  {Kazantzidis} S.,  {Quinn}
  T.~R.,   {Wadsley} J.,  2015, \mn@doi [\mnras] {10.1093/mnras/stv312}, \href
  {https://ui.adsabs.harvard.edu/abs/2015MNRAS.449..494R} {449, 494}

\bibitem[\protect\citeauthoryear{{Ryu}, {Tanaka}, {Perna}  \& {Haiman}}{{Ryu}
  et~al.}{2016}]{Ryu2016}
{Ryu} T.,  {Tanaka} T.~L.,  {Perna} R.,   {Haiman} Z.,  2016, \mn@doi [\mnras]
  {10.1093/mnras/stw1241}, \href
  {https://ui.adsabs.harvard.edu/abs/2016MNRAS.460.4122R} {460, 4122}

\bibitem[\protect\citeauthoryear{{Souza Lima}, {Mayer}, {Capelo}  \&
  {Bellovary}}{{Souza Lima} et~al.}{2017}]{Souza-Lima2017}
{Souza Lima} R.,  {Mayer} L.,  {Capelo} P.~R.,   {Bellovary} J.~M.,  2017,
  \mn@doi [\apj] {10.3847/1538-4357/aa5d19}, \href
  {https://ui.adsabs.harvard.edu/abs/2017ApJ...838...13S} {838, 13}

\bibitem[\protect\citeauthoryear{{Sugimura}, {Hosokawa}, {Yajima}  \&
  {Omukai}}{{Sugimura} et~al.}{2017}]{Sugimura2017}
{Sugimura} K.,  {Hosokawa} T.,  {Yajima} H.,   {Omukai} K.,  2017, \mn@doi
  [\mnras] {10.1093/mnras/stx769}, \href
  {https://ui.adsabs.harvard.edu/abs/2017MNRAS.469...62S} {469, 62}

\bibitem[\protect\citeauthoryear{{Sugimura}, {Hosokawa}, {Yajima}, {Inayoshi}
  \& {Omukai}}{{Sugimura} et~al.}{2018}]{Sugimura2018}
{Sugimura} K.,  {Hosokawa} T.,  {Yajima} H.,  {Inayoshi} K.,   {Omukai} K.,
  2018, \mn@doi [\mnras] {10.1093/mnras/sty1298}, \href
  {https://ui.adsabs.harvard.edu/abs/2018MNRAS.478.3961S} {478, 3961}

\bibitem[\protect\citeauthoryear{{Tagawa}, {Umemura}  \& {Gouda}}{{Tagawa}
  et~al.}{2016}]{Tagawa2016}
{Tagawa} H.,  {Umemura} M.,   {Gouda} N.,  2016, \mn@doi [\mnras]
  {10.1093/mnras/stw1877}, \href
  {https://ui.adsabs.harvard.edu/abs/2016MNRAS.462.3812T} {462, 3812}

\bibitem[\protect\citeauthoryear{{Takeo}, {Inayoshi}, {Ohsuga}, {Takahashi}  \&
  {Mineshige}}{{Takeo} et~al.}{2018}]{Takeo2018}
{Takeo} E.,  {Inayoshi} K.,  {Ohsuga} K.,  {Takahashi} H.~R.,   {Mineshige} S.,
   2018, \mn@doi [\mnras] {10.1093/mnras/sty264}, \href
  {https://ui.adsabs.harvard.edu/abs/2018MNRAS.476..673T} {476, 673}

\bibitem[\protect\citeauthoryear{{Takeo}, {Inayoshi}, {Ohsuga}, {Takahashi}  \&
  {Mineshige}}{{Takeo} et~al.}{2019}]{Takeo2019}
{Takeo} E.,  {Inayoshi} K.,  {Ohsuga} K.,  {Takahashi} H.~R.,   {Mineshige} S.,
   2019, \mn@doi [\mnras] {10.1093/mnras/stz1899}, \href
  {https://ui.adsabs.harvard.edu/abs/2019MNRAS.488.2689T} {488, 2689}

\bibitem[\protect\citeauthoryear{{Toyouchi}, {Hosokawa}, {Sugimura}, {Nakatani}
   \& {Kuiper}}{{Toyouchi} et~al.}{2019}]{Toyouchi2019}
{Toyouchi} D.,  {Hosokawa} T.,  {Sugimura} K.,  {Nakatani} R.,   {Kuiper} R.,
  2019, \mn@doi [\mnras] {10.1093/mnras/sty3012}, \href
  {https://ui.adsabs.harvard.edu/abs/2019MNRAS.483.2031T} {483, 2031}

\bibitem[\protect\citeauthoryear{{Venemans} et~al.,}{{Venemans}
  et~al.}{2013}]{Venemans2013}
{Venemans} B.~P.,  et~al., 2013, \mn@doi [\apj] {10.1088/0004-637X/779/1/24},
  \href {https://ui.adsabs.harvard.edu/abs/2013ApJ...779...24V} {779, 24}

\bibitem[\protect\citeauthoryear{{Viti} et~al.,}{{Viti}
  et~al.}{2014}]{Viti2014}
{Viti} S.,  et~al., 2014, \mn@doi [\aap] {10.1051/0004-6361/201424116}, \href
  {https://ui.adsabs.harvard.edu/abs/2014A&A...570A..28V} {570, A28}

\bibitem[\protect\citeauthoryear{{Watarai}, {Fukue}, {Takeuchi}  \&
  {Mineshige}}{{Watarai} et~al.}{2000}]{Watarai2000}
{Watarai} K.-y.,  {Fukue} J.,  {Takeuchi} M.,   {Mineshige} S.,  2000, \mn@doi
  [\pasj] {10.1093/pasj/52.1.133}, \href
  {https://ui.adsabs.harvard.edu/abs/2000PASJ...52..133W} {52, 133}

\bibitem[\protect\citeauthoryear{{Weingartner} \& {Draine}}{{Weingartner} \&
  {Draine}}{2001}]{Weingartner2001}
{Weingartner} J.~C.,  {Draine} B.~T.,  2001, \mn@doi [\apj] {10.1086/318651},
  \href {https://ui.adsabs.harvard.edu/abs/2001ApJ...548..296W} {548, 296}

\bibitem[\protect\citeauthoryear{{Willott} et~al.,}{{Willott}
  et~al.}{2010}]{Willott2010}
{Willott} C.~J.,  et~al., 2010, \mn@doi [\aj] {10.1088/0004-6256/139/3/906},
  \href {https://ui.adsabs.harvard.edu/abs/2010AJ....139..906W} {139, 906}

\bibitem[\protect\citeauthoryear{{Wu} et~al.,}{{Wu} et~al.}{2015}]{Wu2015}
{Wu} X.-B.,  et~al., 2015, \mn@doi [\nat] {10.1038/nature14241}, \href
  {https://ui.adsabs.harvard.edu/abs/2015Natur.518..512W} {518, 512}

\bibitem[\protect\citeauthoryear{{Yajima}, {Ricotti}, {Park}  \&
  {Sugimura}}{{Yajima} et~al.}{2017}]{Yajima2017}
{Yajima} H.,  {Ricotti} M.,  {Park} K.,   {Sugimura} K.,  2017, \mn@doi [\apj]
  {10.3847/1538-4357/aa8269}, \href
  {https://ui.adsabs.harvard.edu/abs/2017ApJ...846....3Y} {846, 3}

\bibitem[\protect\citeauthoryear{{Yan}, {Sadeghpour}  \& {Dalgarno}}{{Yan}
  et~al.}{1998}]{Yan1998}
{Yan} M.,  {Sadeghpour} H.~R.,   {Dalgarno} A.,  1998, \mn@doi [\apj]
  {10.1086/305420}, \href
  {https://ui.adsabs.harvard.edu/abs/1998ApJ...496.1044Y} {496, 1044}

\bibitem[\protect\citeauthoryear{{Zeilig-Hess}, {Levinson}  \&
  {Nakar}}{{Zeilig-Hess} et~al.}{2019}]{Zeilig-Hess2019}
{Zeilig-Hess} M.,  {Levinson} A.,   {Nakar} E.,  2019, \mn@doi [\mnras]
  {10.1093/mnras/sty3034}, \href
  {https://ui.adsabs.harvard.edu/abs/2019MNRAS.482.4642Z} {482, 4642}

\bibitem[\protect\citeauthoryear{{eLISA Consortium} et~al.,}{{eLISA Consortium}
  et~al.}{2013}]{LISA2013}
{eLISA Consortium} et~al., 2013, arXiv e-prints, \href
  {https://ui.adsabs.harvard.edu/abs/2013arXiv1305.5720E} {p. arXiv:1305.5720}

\makeatother
\end{thebibliography}

\bsp	
\label{lastpage}
\end{document}